\documentclass[final,5p,times,twocolumn]{elsarticle}

\usepackage{graphics}
\usepackage{graphicx}
 \usepackage{epsfig}

\usepackage{amssymb}
\usepackage{amsmath}
\usepackage{hyperref}

\usepackage{lineno}
\newcommand*\patchAmsMathEnvironmentForLineno[1]{%
  \expandafter\let\csname old#1\expandafter\endcsname\csname #1\endcsname
  \expandafter\let\csname oldend#1\expandafter\endcsname\csname end#1\endcsname
  \renewenvironment{#1}%
     {\linenomath\csname old#1\endcsname}%
     {\csname oldend#1\endcsname\endlinenomath}}%
\newcommand*\patchBothAmsMathEnvironmentsForLineno[1]{%
  \patchAmsMathEnvironmentForLineno{#1}%
  \patchAmsMathEnvironmentForLineno{#1*}}%
\AtBeginDocument{%
\patchBothAmsMathEnvironmentsForLineno{equation}%
\patchBothAmsMathEnvironmentsForLineno{align}%
\patchBothAmsMathEnvironmentsForLineno{flalign}%
\patchBothAmsMathEnvironmentsForLineno{alignat}%
\patchBothAmsMathEnvironmentsForLineno{gather}%
\patchBothAmsMathEnvironmentsForLineno{multline}%
}
\renewcommand{\vec}[1]{\mathbf{#1}}
\usepackage{color}

\journal{Computer Physics Communications}

\begin{document}

\begin{frontmatter}



\title{Elimination of the linearization error and improved basis-set
convergence within the FLAPW method}


\author{Gregor Michalicek\corref{cor1}}
\ead{g.michalicek@fz-juelich.de}
\author{Markus Betzinger\corref{cor2}}
%\ead{m.betzinger@fz-juelich.de}
\author{Christoph Friedrich\corref{cor2}}
%\ead{c.friedrich@fz-juelich.de}
\author{Stefan Bl\"ugel\corref{cor2}}
%\ead{s.bluegel@fz-juelich.de}

\cortext[cor1]{Corresponding author.}

\address{Peter Gr\"unberg Institut and Institute for Advanced Simulation, Forschungszentrum J\"ulich and JARA, D-52425 J\"ulich, Germany}

\begin{abstract}
We analyze in detail the error that arises from the linearization in
linearized augmented-plane-wave (LAPW) basis functions around predetermined energies
$E_l$ and show that it can lead to undesirable dependences of the calculated results on 
method-inherent parameters such as energy parameters $E_l$ and muffin-tin sphere radii. 
To overcome these dependences, we evaluate approaches 
that eliminate the linearization error systematically by adding local orbitals 
(LOs) to the basis set. We consider two kinds of LOs: (i) constructed from 
solutions $u_l(r,E)$ to the scalar-relativistic approximation of the radial Dirac equation with $E>E_l$ and (ii)
constructed from second energy derivatives 
$\partial^2 u_l(r,E) / \partial E^2$ at $E=E_l$.
We find that the latter eliminates the error most efficiently and yields the density functional answer to many
electronic and materials properties with very high precision.
Finally, we demonstrate that the so constructed LAPW+LO basis shows a more favorable convergence behavior than the conventional LAPW basis due to a better decoupling of muffin-tin and interstitial regions, similarly to the related APW+lo approach, which requires an extra set of LOs to reach the same total energy, though.
\end{abstract}


\end{frontmatter}

% \linenumbers


\section{Introduction}
\label{introduction}

In the past decades, material simulations have become an invaluable approach in 
condensed matter physics and materials science. Ever increasing computer power as well as theoretical 
and methodological progress in the description of materials are the main
incentives for more and more accurate calculations on more and more complex 
materials. Within the wide range of theoretical approaches, density
functional theory (DFT)~\cite{PhysRev.136.B864} is the method 
of choice for the calculation of electronic ground-state properties of materials. 

Practical realizations of DFT almost invariably rely on the Kohn-Sham (KS) formalism~\cite{PhysRev.140.A1133}, 
which employs an auxiliary system of noninteracting electrons whose number density
coincides with that of the real interacting system. Most codes make use of a
set of basis functions to represent the quantum mechanical wave functions
of these noninteracting electrons, which enables a formulation of the underlying
differential KS equation as a generalized eigenvalue problem. In recent years
we witnessed a trend toward the investigation of solids of increasing electronic, chemical, and 
structural complexity, solids that exhibit narrow electronic bands, large band gaps, electrons
that contribute to physisorption and chemisorption, which are in turn described by more sophisticated
exchange and correlation functionals, e.g.,~hybrid functionals, the exact-exchange functional in the 
optimized-effective-potential method, van der Waals functionals, or correlation functionals based on 
the random phase approximation (RPA), just to name a few. The description of the electronic structure and the application
of the new functionals raise new challenges to the efficiency and ability of the basis set to precisely represent
the density-functional answer. The aim of this paper is to evaluate and improve the LAPW basis for this purpose.

The simplest basis set for systems with periodic boundary conditions is certainly the plane-wave basis.
The accuracy of which is controlled by a single convergence
parameter, the momentum cutoff radius $G_\mathrm{max}$. However, the rapid variations of the
wave functions close to the atomic nuclei cannot be resolved in practice
with this basis,
and one has to resort to pseudopotentials and pseudized wave functions
within a certain distance from the atomic nuclei~\cite{Pickett1989115}. The core electrons are then incorporated into the
pseudopotentials, and only the valence electrons are treated explicitly. 

The pseudopotential approximation effectively restricts the range of materials
that can be examined. Compounds containing 4$f$ and 5$f$ elements, and transition-metals as well as
oxides, nitrides, and carbides cannot be treated efficiently within this approach. Among the all-electron
approaches that describe core and valence electrons on an equal footing
(Gaussian functions~\cite{Gaussian-1}, the projector augmented-wave~\cite{PhysRevB.50.17953} 
and the linearized muffin-tin orbitals
method~\cite{PhysRevB.12.3060,skriver-book,MSC-LMTO-inbook} to name a few),
the full-potential linearized augmented-plane-wave (FLAPW) 
method~\cite{PhysRevB.12.3060,0305-4608-5-11-016,PhysRevB.24.864} 
provides one of the most precise basis sets for all-electron calculations.
It allows for studying the electronic structure of a large variety of materials,
including open systems with low symmetry and compounds of any chemical composition.

The FLAPW method is based on a partitioning of space into 
non-overlapping spheres centered at the atomic nuclei, 
the so-called muffin-tin (MT) spheres, and the interstitial region (IR). 
The core states are completely confined within the spheres, which allows 
to treat them as localized states in a spherically symmetric atomic potential. 
For the valence electrons, on the other hand, the APW basis functions are defined piecewise:
plane waves of all reciprocal lattice vectors up to the maximal momentum $G_\textrm{max}$ in the IR,
which are augmented by radial functions in the MT spheres that are solutions of the 
scalar-relativistic approximation to the Dirac equation for the spherically averaged effective potential.
Employing the linearized APW (LAPW) basis set, the energy-dependent radial functions are approximated by 
energy-independent functions evaluated
at predetermined energy parameters $E_l$. The functions are linearly combined 
so as to match to the plane waves in value and slope at the MT sphere
boundaries. In the conventional LAPW basis
there are two radial functions per angular momentum quantum number $l$, the
solution $u_l(r,E_l)$ and its energy derivative $\dot{u}_l(r,E_l)=\partial
u_l(r,E)/\partial E|_{E=E_l}$. In this way, radial functions $u_l(r,E)$ with $E$ close to $E_l$ can be described up to linear order in $E-E_l$.

This conventional LAPW basis
set is a very accurate one for a wide range of materials. In comparison to
the pure plane-wave basis, the LAPW basis requires far less basis functions, while
still being able to provide an all-electron description.
However, no matter how large the momentum cutoff radius is chosen,  the flexibility of
the basis in the MT spheres is restricted to the two radial functions 
$u_l(r,E_l)$ and $\dot{u}_l(r,E_l)$ per $l$
quantum number and the energy-dependent radial function is approximated by 
these energy-independent functions. This gives rise to a linearization error that can
notably affect the accuracy, for example, for materials with large 
bandwidths, large bandgaps, or if states are considered that are energetically far away from 
the energy parameters \cite{AccWaveLAPW,PhysRevB.74.045104}.
It is obvious that the linearization error
depends on two sets of method-inherent parameters, (i) the energy parameters $E_l$ 
and (ii) the MT radii, $R_\textrm{MT}$, as they define the region of space in which the wave 
function is represented by $u_l(r,E_l)$ and $\dot{u}_l(r,E_l)$. None of the parameters can 
be chosen on the basis of a variational principle. Optimally, the final
results should not depend on $E_l$ and $R_{\mathrm{MT}}$ as long as they are chosen
within a reasonable range of values. 

Over time, several approaches have been proposed to reduce the linearization 
error. The first approach that we mention here is 
the quadratic APW (QAPW) method~\cite{QuadraticAPW-2,QuadraticAPW}. 
In this method the MT augmentation is extended by including 
the second-order energy derivative $\ddot{u}_l(r,E_l)$ and employing
an algebraic relation for
the matching coefficients of $\dot{u}_l(r,E_l)$ and $\ddot{u}_l(r,E_l)$.

Next, Singh~\cite{PhysRevB.43.6388,SinghBook} investigated how to deal with 
semicore states. These are high-lying core states that are not completely 
confined within the MT spheres and therefore cannot be treated separately 
from the valence states. He introduced the radial functions $u_l(r,E^\mathrm{SC}_l)$ 
that solve the scalar-relativistic Dirac equation with 
energy parameters $E^\mathrm{SC}_l \ll E_l$ that correspond 
to the energy of the respective semicore state. He introduced and compared the inclusion of 
these functions by either enforcing continuity of the basis-function curvature as 
additional matching condition at the MT boundaries or constructing additional local 
orbitals (LOs) with $u_l(r,E_l)$, $\dot{u}_l(r,E_l)$, and $u_l(r,E^\mathrm{SC}_l)$ 
that are nonzero only in the MT spheres. He found that the first approach is effective
in describing the semicore states, on the expense of creating stiffer basis functions
across the MT boundary imposed by the additional matching conditions. 
As a result, the basis
set becomes less flexible and larger basis sets are required to achieve the same accuracy.
On the other hand, the extension of the basis with LOs avoids this problem.

The extended LAPW (ELAPW) basis developed by Krasovskii 
\textit{et al.}~\cite{ELAPW-1,ELAPW-kp,PhysRevB.56.12866} is another approach
amending the conventional LAPW basis set by LOs
to improve the description of the electronic structure over a wide energy window. 
Here, pairs of LOs are constructed from $u_l(r,E^\mathrm{LO}_l)$ and 
$\dot{u}_l(r,E^\mathrm{LO}_l)$, respectively, at energy 
parameters $E^\mathrm{LO}_l$ that are typically placed at energies above the Fermi energy.
However, the authors did not give a prescription of how to choose them
in an optimal way. Friedrich \textit{et al.}~\cite{PhysRevB.74.045104} considered 
LOs that are constructed from second-order energy
derivatives $\ddot{u}_l(r,E_l)$ taken at the valence energy parameters $E_l$
to improve the description of high-lying states for $GW$ calculations.
Betzinger \textit{et al.}~\cite{PhysRevB.83.045105} employed LOs defined at higher energies to refine the susceptibility matrix
in optimized-effective-potential calculations.

Sj\"ostedt \textit{et al.}~\cite{Sjoestedt200015} used LOs to achieve a linearization of the 
basis set, known as 
the APW+lo method, in which the condition of continuous slope across the MT
sphere boundary is dropped for the basis functions. The functions $\dot{u}_l(r,E_l)$ are not used anymore as
augmentation functions but instead are included in LOs.
This makes the basis set
more flexible when compared with the FLAPW method. Thus, the same accuracy can be achieved
with less basis functions and this saves computation time. We will show later that
the inclusion of LOs, in particular second-derivative LOs, into the LAPW
basis can yield a similar
gain in flexibility.

In this article we evaluate the conventional LAPW basis set for a set of systems exhibiting large muffin-tin radii 
and large band gaps. We have chosen fcc Ce, KCl in the rock-salt structure, fcc Ar and bcc V as test systems. We examine
the linearization error and find that physical quantities such as the total energies, lattice constants, or 
band structures are susceptible to slight variations of method-inherent parameters such as MT radii and energy parameters. 
We apply two types of local orbitals to extend the LAPW basis set, (i) one based on functions of higher energy derivative (LAPW+HDLO) and (ii) one based on energy parameters at higher energy (LAPW+HELO). We show that these LOs are very effective in 
reducing or, within the relevant degree of accuracy, even in eliminating the linearization error. Here, the 
second-derivative HDLOs are particularly efficient.

The article continues with a brief introduction to the LAPW basis and the definition of the LO extensions in 
Sec.~\ref{lapwBasisSection}. In Sec.~\ref{materialsAndParameters}, we introduce the test 
set of materials for which our evaluations are carried out. We continue in Sec.~\ref{theLinearizationError}
with a discussion of the linearization error for the conventional and the LO extended LAPW basis sets. 
In detail, in Sec.~\ref{representationError} we evaluate the quality of the various basis sets in 
their ability to represent Kohn-Sham states inside the MT sphere, 
discuss dependences of computed physical properties on method-inherent parameters 
in Sec.~\ref{calculationDependencesSection}, and investigate the effect of the linearization
error on unoccupied states, in particular, the KS band gap in Sec.~\ref{unoccupiedStates}. 
Finally, we discuss in Sec.~\ref{basisSetSize} the effect of the LOs on the
convergence behavior of the total energy with respect to the basis set size for the different 
basis sets and also in comparison 
with the APW+lo approach, before we conclude in Sec.~\ref{conclusion}.

\section{LAPW basis and LO extensions}
\label{lapwBasisSection}

In the KS formalism of DFT~\cite{PhysRev.140.A1133} one considers a fictitious system of noninteracting
electrons moving under the influence of an effective potential
$V^\mathrm{eff}(\vec r)$. For lattice periodic potentials the electronic quantum mechanical wave functions
$\varphi_{\vec{k}n}(\vec r)$ with the Bloch vector $\vec k$ and the band
index $n$ are thus solutions of
\begin{equation}
\left[-\frac{1}{2}\Delta+V^\mathrm{eff}(\vec{r})\right]\varphi_{\vec{k}n}(\vec r)
=\epsilon_{\vec{k}n}\varphi_{\vec{k}n}(\vec r)\,,
\label{KSeq}
\end{equation}
where the spin index has been suppressed. 
The effective potential is defined such that the KS
system exhibits the same number density as that of the real interacting
system. It 
comprises the electrostatic potential created by the charged particles --
electrons and nuclei -- as well as an exchange-correlation potential, for
which we use here the Perdew-Zunger parametrization of the local-density 
approximation \cite{PhysRevB.23.5048}. 
For simplicity, we present in Eq.~(\ref{KSeq}) and in the following the
non-relativistic equations, while in practice the scalar- or
fully relativistic Dirac equation is employed. 

Close to the atomic nuclei the effective potential is predominantly
spherical, which allows the core states to be determined efficiently from the
fully relativistic radial Dirac equation. The core states fall off quickly
toward the MT sphere boundary, where they are practically zero. It can be
shown~\cite{PhysRevB.74.045104,SinghBook} that this property guarantees that the core states are
orthogonal to the radial functions used for the construction of the 
LAPW basis for the valence electrons, which we introduce in the following.
The valence states are represented by the set of basis functions
\begin{equation}
\phi_{\vec{k}\vec{G}}(\vec{r}) = 
\left\{ \begin{array}{l l}  \frac{1}{\sqrt{\Omega}} e^{i(\vec{k} + \vec{G})\vec{r}} & \text{for } \vec{r} \in \text{IR} \\ 
\sum\limits_{L} R_{L\alpha}^{\vec{k}\vec{G}}(r_\alpha,E_{l\alpha})
Y_L(\vec{\hat{r}}_\alpha) & \text{for } \vec{r} \in \text{MT$_\alpha$}\\ \end{array}\right.,
\label{lapwBasisEquation}
\end{equation}
where the $\vec{G}$ are reciprocal lattice vectors, $\Omega$ is the unit-cell
volume, $L=(l,m)$ comprises the angular momentum and the magnetic quantum
number, $\vec{r}_\alpha$ is the position vector relative to the MT sphere center $\vec{R}^\alpha$ of atom
$\alpha$, and $Y_L(\hat{\vec{r}})$ are the spherical harmonics. In practice,
one employs the cutoff parameter $G_\mathrm{max}$ that controls the number of basis functions 
through $|\vec{k+G}|\le
G_\mathrm{max}$ and implies the cutoff parameter $l_{\mathrm{max}\alpha}$ with $l\le l_{\mathrm{max}\alpha}$ through the rule of thumb $l_{\mathrm{max}\alpha}=G_\mathrm{max}R_{\mathrm{MT}\alpha}$ \cite{SinghBook}.

For each angular momentum quantum number $l$, the radial functions
\begin{equation}
R_{L\alpha}^{\vec{k}\vec{G}}(r_\alpha,E_{l\alpha}) = a_{L\alpha}^{\vec{k}\vec{G}} u_{l\alpha}(r_\alpha,E_{l\alpha}) + b_{L\alpha}^{\vec{k}\vec{G}} \dot{u}_{l\alpha}(r_\alpha,E_{l\alpha})
\label{Rfunc}
\end{equation}
are linear combinations of the normalized solution to
\begin{equation}
\left\{-\frac{1}{2}\frac{\partial^2}{\partial r^2}+\frac{l(l+1)}{2r^2}+V^\mathrm{eff}_{\alpha}(r)\right\}
r\,u_{l\alpha}(r,E)=E\,r\,u_{l\alpha}(r,E)\,
\label{udiff}
\end{equation}
for $r \in \mathrm{MT}_\alpha$ and its energy derivative $\dot{u}_{l\alpha}(r,E_{l\alpha})$, where
$E=E_{l\alpha}$ and $V^\mathrm{eff}_{\alpha}(r)$ is the spherical part of 
$V^\mathrm{eff}(\vec{r})$ inside of $\mathrm{MT}_\alpha$ 
and expressed in Hartree units (1~Htr = 2~Ry $\approx$ 27.2~eV).
The matching coefficients $a_{L\alpha}^{\vec{kG}}$ and $b_{L\alpha}^{\vec{kG}}$ 
are determined such that the
$\phi_{\vec{k}\vec{G}}(\vec{r})$ are continuous in value and slope at the MT
sphere boundaries.

The energy $E_{l\alpha}$ is a parameter that is fixed for each iteration of
a self-consistent-field run. If it happens to be identical to the KS
eigenvalue $\epsilon_{\vec{k}n}$,
the function $u_{l\alpha}(r,E_{l\alpha})Y_L(\hat{\vec{r}})$ already solves
Eq.~(\ref{KSeq}) pointwise in the MT sphere by construction (provided that we restrict ourselves to
the spherical part of the effective potential, which is indeed much larger
than the nonspherical terms). In this sense, the inclusion of the energy
derivative $\dot{u}_{l\alpha}(r,E_{l\alpha})$ makes it possible that
states with energies different from but close to $E_{l\alpha}$ can also be
described accurately, in accordance with the Taylor expansion to linear order
\begin{equation}
u_{l\alpha}(r,E) \approx 
u_{l\alpha}(r,E_{l\alpha})+(E-E_{l\alpha})\dot{u}_{l\alpha}(r,E_{l\alpha})\,.
\label{Taylor}
\end{equation}

The energy parameters $E_{l\alpha}$ should be chosen as close as possible to the 
corresponding band energies. While the bands of a crystal exhibit a strong dependence 
on the Bloch vector $\vec{k}$ and the band index $n$, the energy parameters only 
depend on the atom and the angular momentum $l$. Thus, differences between the band 
energies and the energy parameters are unavoidable.

In practice, several methods are in use to choose the energy parameters automatically
in each iteration of the self-consistent-field cycle. One example
for such a method solves for each sphere an atomic Hamiltonian employing
the spherical part of the effective KS potential in the MT sphere with a 
confining potential outside. The energy parameters of the valence states 
are then set to the corresponding atomic eigenenergies. The so-chosen energy 
parameter, that we denote as \textit{atomic energy parameter}, is 
motivated by the fact that the bands in a solid form out of the atomic states.
However, this approach may yield energy parameters above the Fermi energy since it 
does not take the occupation of states into account. Energy parameters below the Fermi 
energy are obtained by another method that
sets them at the \textit{energy center of mass} of the $l$-resolved partial density 
of the occupied valence states.
The disadvantage of the latter method is that systems with narrow bands at the Fermi energy,
where changes in the occupation of bands can occur easily between
two subsequent self-consistent-field iterations, exhibit a stronger variation of the energy parameters,
which at the end may reduce the speed to self-consistency or even make the self-consistency process less stable than the
former choice of $E_l$, which is less sensitive to these band reorderings. Both methods also depend on the chosen MT radii.\footnote{For 
completeness we note that we employ the atomic energy parameters for $E_l$ if not 
stated otherwise. The qualitative results do not depend on the particular choice of the parameters.}

Of course, the more the KS eigenvalues differ from the energy parameter, the
less adequate the basis becomes for the corresponding KS eigenfunctions, 
which we refer to as the linearization error.
From Eq.~(\ref{Taylor}) a solution seems obvious: one simply adds the
second-order derivative $\ddot{u}_{l\alpha}(r,E_{l\alpha})$ to
Eq.~(\ref{Rfunc}). However, this would require an additional expansion
coefficient and, as a consequence, an additional matching condition at the
MT boundaries. It was shown~\cite{PhysRevB.43.6388} that this leads to a less flexible
(slower convergent) basis set.
The LO construction~\cite{PhysRevB.43.6388} avoids this problem.

LOs are additional basis functions
\begin{equation}
\phi_{L\alpha}^{\mathrm{LO}}(\vec{r}) = R_{l\alpha}^{\mathrm{LO}}(r_\alpha) Y_L(\vec{\hat{r}}_\alpha)
\end{equation}
that are completely confined to a MT sphere. There are $2l+1$ LOs per $l$
quantum number. Their radial part is a linear
combination
\begin{equation}
R_{l\alpha}^{\mathrm{LO}}(r_\alpha) = 
a_{l\alpha}^{\mathrm{LO}} u_{l\alpha}(r_\alpha,E_{l\alpha}) + 
b_{l\alpha}^{\mathrm{LO}} \dot{u}_{l\alpha}(r_\alpha,E_{l\alpha}) + 
c_{l\alpha}^{\mathrm{LO}} u_{l\alpha}^{\mathrm{LO}}(r_\alpha)\, ,
\end{equation}
where the coefficients are determined by enforcing normalization 
$\int r_\alpha^2 R^\mathrm{LO}_{l\alpha}(r_\alpha)^2dr_\alpha=1$
as well as zero value and slope at the MT boundary, implying that the LOs are zero in the IR.
The third radial function $u_{l\alpha}^\mathrm{LO}(r_\alpha)$ can be chosen
rather arbitrarily. For a semicore state it should be the solution of 
Eq.~(\ref{udiff}) with the eigenenergy of the semicore state as the energy 
parameter for the given angular momentum $l$~\cite{PhysRevB.43.6388}. The conduction-band spectrum may be improved
by adding LOs with energy parameters chosen in the corresponding
energy range~\cite{PhysRevB.83.045105,PhysRevB.83.081101}. The third radial function can also be taken to be the 
second-order energy derivative $\ddot{u}_{l\alpha}(r,E_{l\alpha})$, which
would correspond to the next higher term in Eq.~(\ref{Taylor})~\cite{PhysRevB.74.045104}.

\begin{figure}
\centering
\begin{raggedright}(a)\par\end{raggedright}
\includegraphics*[width=\columnwidth]{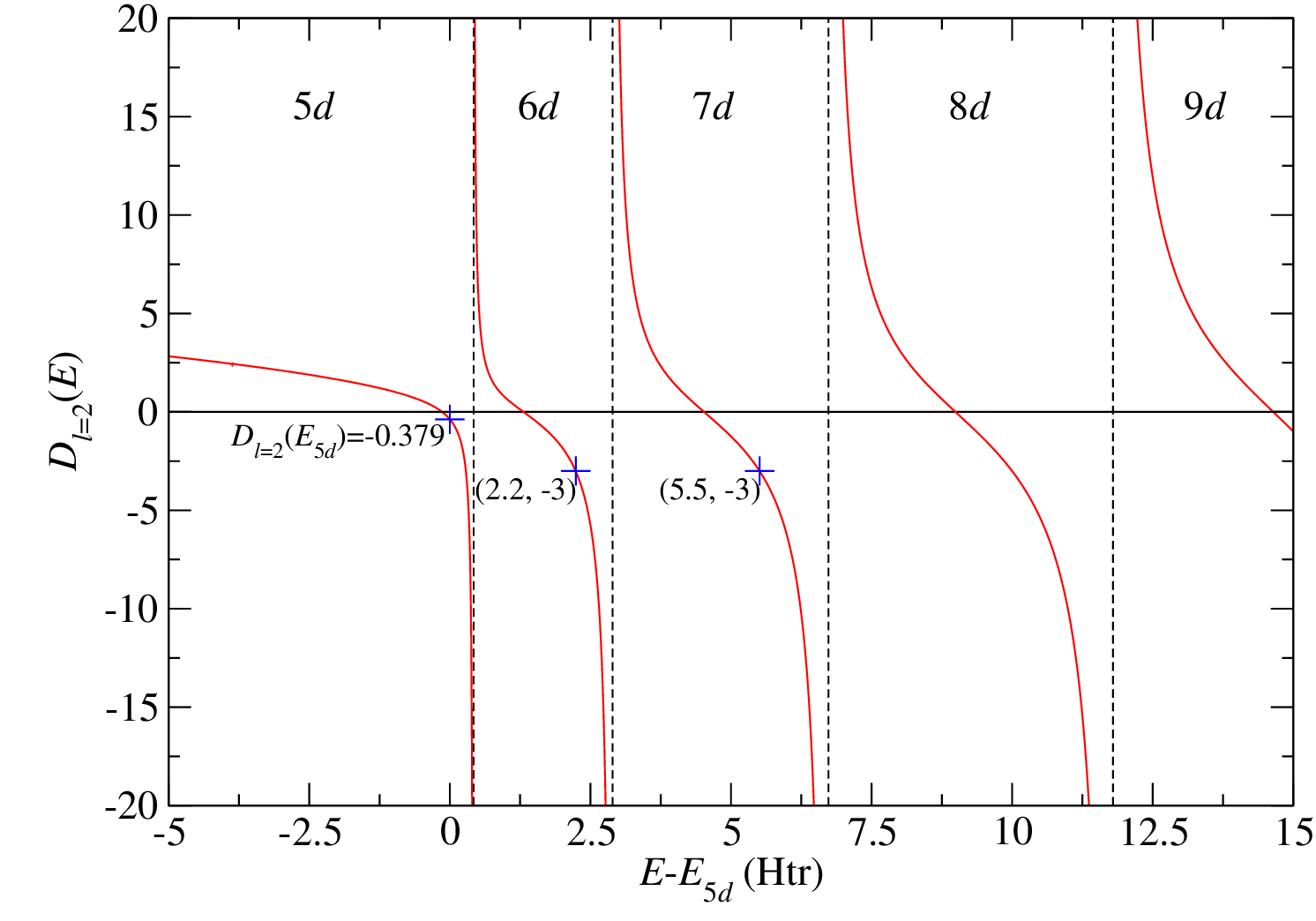}
\begin{raggedright}(b)\par\end{raggedright}
\includegraphics*[width=\columnwidth]{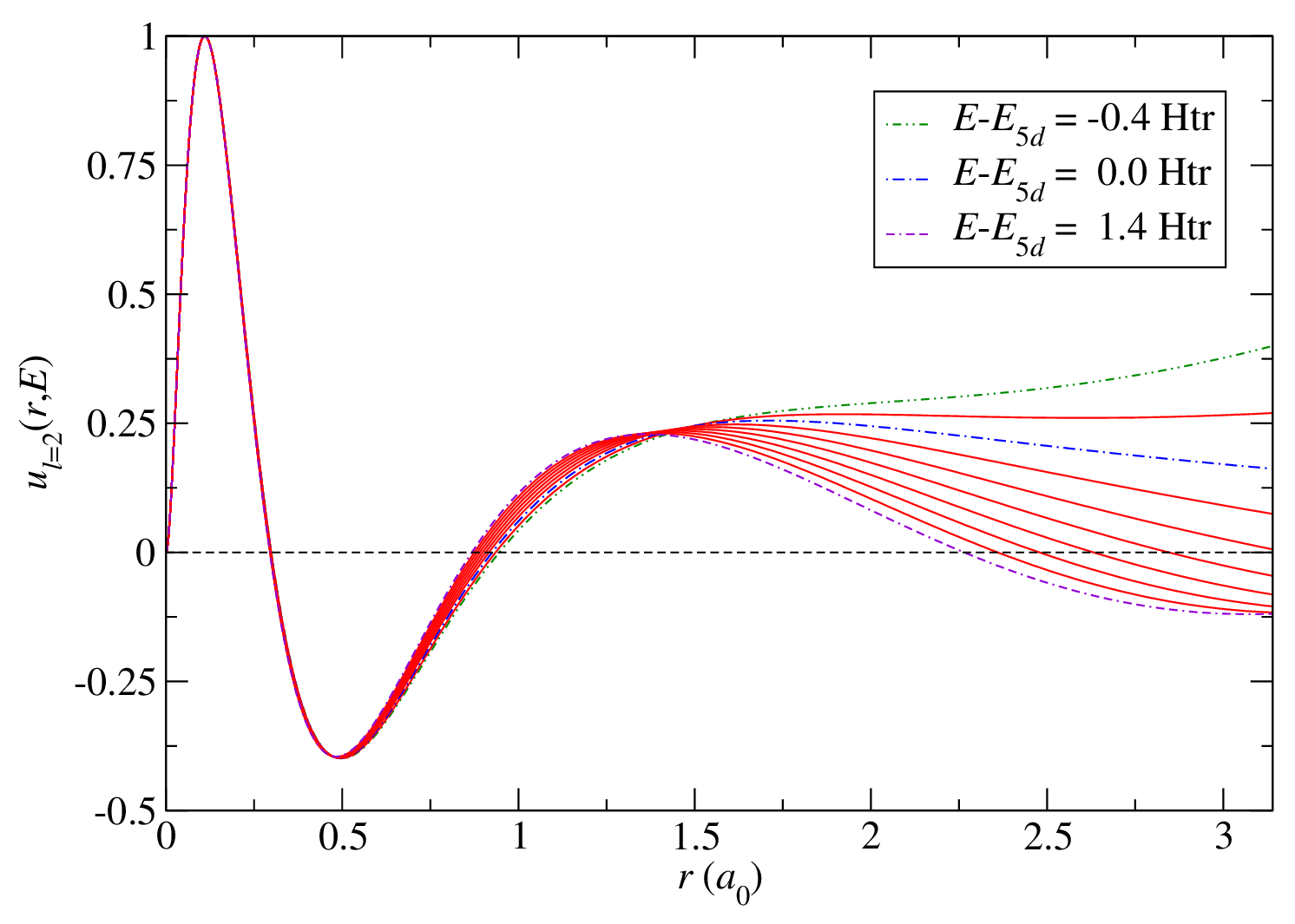}
\caption{(a) Logarithmic derivative $D_l(E)$ for the $d$ channel of fcc Ce as a function 
of the energy relative to the $5d$ energy parameter $E_{5d}$.
The energy parameters for the two additional sets of LOs in the
LAPW+HELO$\times$1 and LAPW+HELO$\times$2 basis sets are marked. (b) Solutions of
Eq.~(\ref{udiff}) obtained for different energies $E$. The functions are not
normalized but scaled to coincide at the first maximum. The energy
dependence increases towards the MT sphere boundary at $3.14~a_0$.}
\label{logDerivCeFigure}
\end{figure}

We consider here two kinds of LOs: (i) Higher-derivative LOs (HDLOs), where the
third radial function $u_{l\alpha}^\mathrm{LO}(r)$ is set equal to
$\ddot{u}_{l\alpha}(r,E_{l\alpha})$ (here we only consider the second
derivative) and (ii) Higher-energy LOs (HELOs), where
$u_{l\alpha}^\mathrm{LO}(r )=u_{l\alpha}(r,E^\mathrm{LO}_{l\alpha})$ with 
$E_{l\alpha}^\mathrm{LO}>E_{l\alpha}$. The functions $\ddot{u}_{l\alpha}(r,E_{l\alpha})$
for the HDLOs are obtained by solving the second energy derivative of Eq.~\eqref{udiff}.
For the HELOs we determine
$E_{l\alpha}^\mathrm{LO}$ by the condition that the logarithmic derivative
\begin{equation}
D_{l\alpha}(E) =
\left.\frac{u'^{\mathrm{LO}}_{l\alpha}(r,E)}{u^{\mathrm{LO}}_{l\alpha}(r,E)}\right|_{r=R_{\mathrm{MT}_\alpha}}
\label{logder}
\end{equation}
is equal to the constant $-(l+1)$, corresponding to the evanescent solution of Eq.~(\ref{udiff}) with
$V^\mathrm{eff}_{\alpha}=E=0$~\cite{PhysRevB.83.045105}. Because of the form of $D_{l\alpha}(E)$
there are infinitely many solutions that are all orthogonal 
to each other and each being related to a particular principal 
quantum number. This allows a systematic extension of the basis. Figure 
\ref{logDerivCeFigure}(a) displays the logarithmic derivatives for the $d$
channel of fcc cerium as an example, where the intersection with $D_l=-3$ gives the set of energy parameters $E^\mathrm{LO}_{l=2}$ for Ce. 
For reference, we plot in Fig.~\ref{logDerivCeFigure}(b) the radial function $u_{l=2}(r,E)$ for
different energies $E$ covering a range of $1.8\,\mathrm{Htr}$.
The pole of $D_l$ at around $E=0.4\,\mathrm{Htr}$
in Fig.~\ref{logDerivCeFigure}(a) marks the transition from $5d$ to $6d$ orbitals and is related to the zero value  
of the corresponding radial function at the sphere boundary, the node at around 
$E=1.3\,\mathrm{Htr}$ in Fig.~\ref{logDerivCeFigure}(a) corresponds to a vanishing slope in (b).

Fig.~\ref{logDerivCeFigure}(b) displays the energy dependence of the Ce 5$d$ wave function, $u_l(r,E)$, inside the MT sphere.
In the vicinity of the atomic nucleus up to a radius of about 1.5~$a_0$, $u_l(r,E)$ is not very sensitive to the particular choice of energy 
$E$, but it is determined by the singularity of the effective potential and the angular momentum barrier. This changes towards 
the MT boundary, where chemical bonding determines the details of the potential, and a strong energy dependence of the basis 
function is displayed. As a consequence, one can expect that the choice of the MT radius may have a significant influence on
the linearization error. With increasing MT radius we expect an increase of the error. On the other hand a larger MT radius is advantageous
since it implies a smaller number of required basis functions or $G_\textrm{max}$, respectively. We mention in passing
that a reduction of the MT radius broadens the branches of the logarithmic derivative. This also shifts the HELO energies upwards.

In this work, we investigate and compare HDLO and HELO extensions 
to the LAPW basis. We work with two different sets of HELOs that include one and two sets of LOs of successive principle numbers,
denoted by HELO$\times$1 and HELO$\times$2, respectively. The LAPW+HDLO$\times$1 and LAPW+HELO$\times$1 
basis sets comprise $16$ additional LOs per atom with $l=0,1,2,3$, while the LAPW+HELO$\times$2
basis contains $32$ additional LOs per atom. 

\section{Investigated materials and calculational parameters}
\label{materialsAndParameters}

The quality of the different basis sets is analyzed by studying the linearization error
and its consequences for calculated physical quantities for a test set of representative materials,
i.e.~materials for which a significant linearization error can be expected. This includes materials
with large MT radii (around 3~$a_0$), large band gaps (between 5-10~eV), or cases where 
the choice of energy parameters deviates significantly from the electronic state to be described. 
In total we have selected fcc Ce, KCl in the rock-salt structure, fcc Ar, and bcc V. Typical 
cut-off parameters such as the number of $\vec{k}$ points in the irreducible wedge of the Brillouin zone
(IBZ) or the number of basis functions per atom controlled by the product $G_\textrm{max}R_\textrm{MT}$
are chosen such that the quantities discussed in the paper are converged with respect to these parameters.
The calculational parameters for the different materials are shown in 
Table~\ref{calculationParametersTable}.

\begin{figure}
\centering
\includegraphics*[width=\columnwidth]{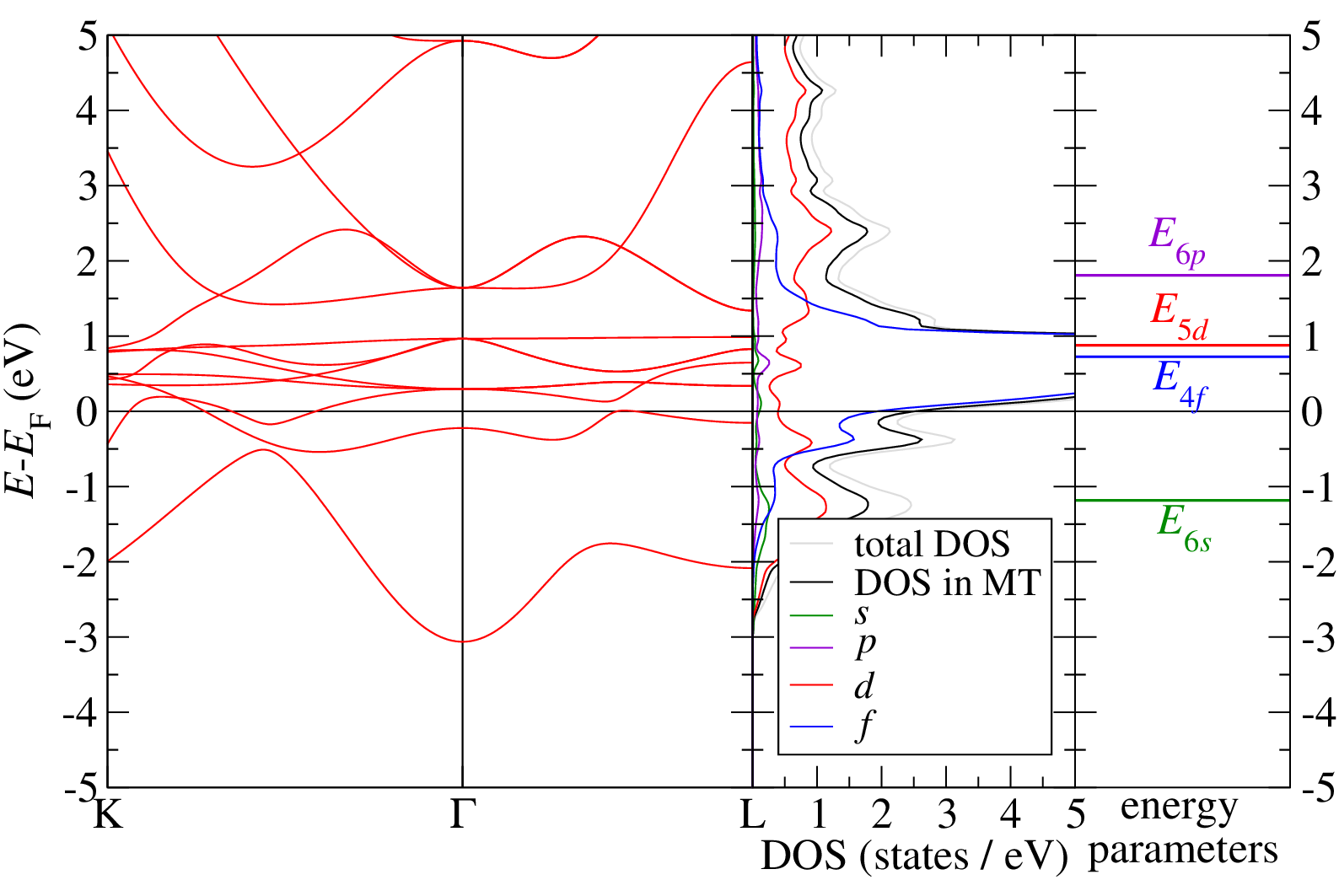}
\caption{Density of states and band structure of fcc Ce. The left panel shows the band structure along the 
high-symmetry lines K$-\Gamma-$L. The center panel shows the associated
total density of states, as well as its projection onto the MT sphere and onto the respective $s$, 
$p$, $d$, and $f$ channels. The right panel displays the atomic energy parameters.}
\label{dosBandsCeFigure}
\end{figure}

\begin{figure}
\centering
\includegraphics*[width=0.8\columnwidth]{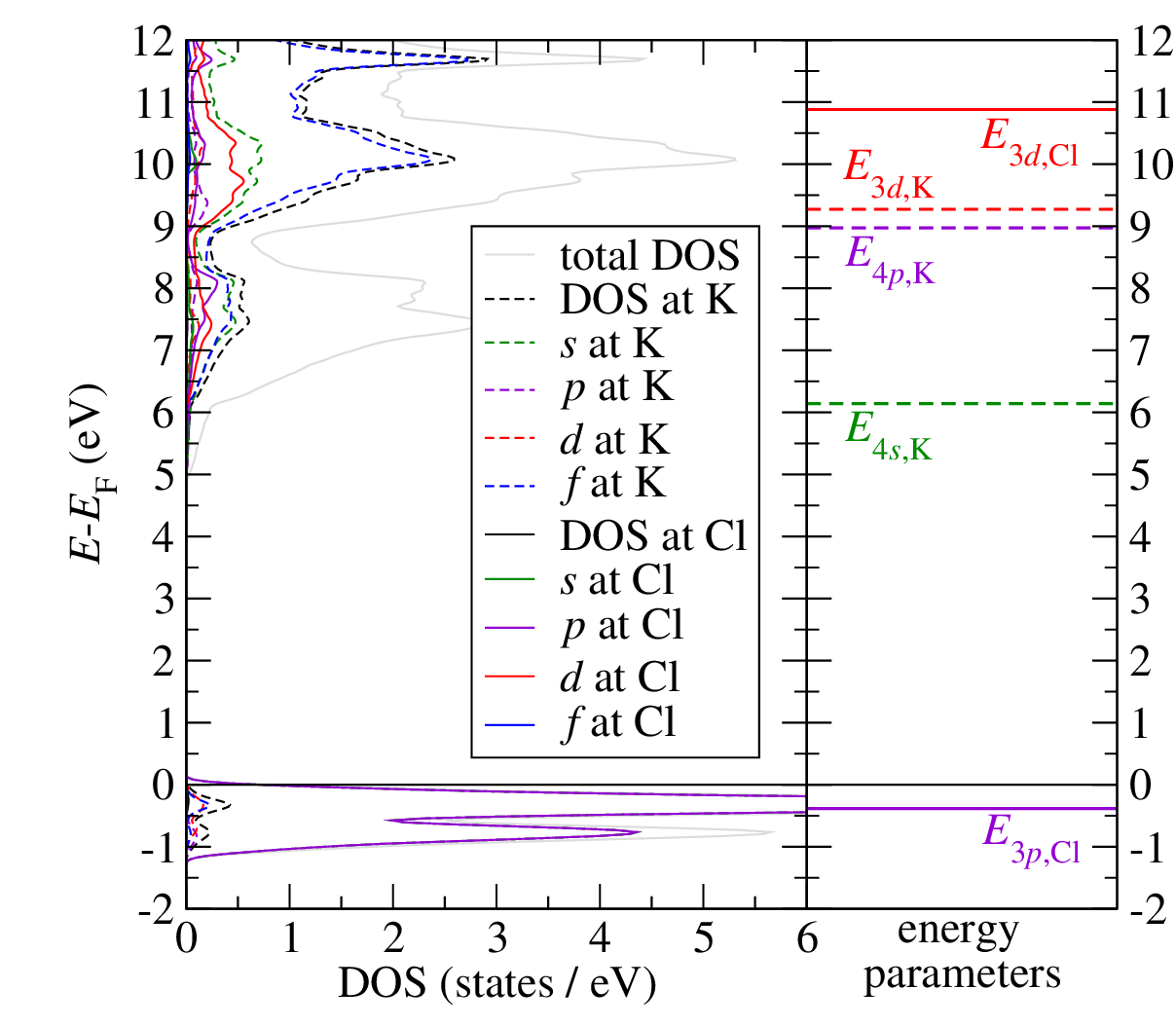}
\caption{Density of States and atomic energy parameters for KCl.}
\label{dosEParamKClFigure}
\end{figure}

The first material to be taken under scrutiny is fcc cerium in a nonmagnetic configuration.
Its choice is largely motivated by previous convergence studies of APW-type basis sets documented in 
literature~\cite{Sjoestedt200015}. The material allows to employ a large MT radius which is 
typically chosen to be larger than $3~a_0$ ($a_0$ is the Bohr radius). As we will see in Sec.~\ref{basisSetSize}, the 
chosen values for the convergence parameters (cf.\ Table~\ref{calculationParametersTable}) lead to 
absolute convergence of the total energy of about 14.4~meV 
for the conventional LAPW basis, of about 3.3~meV for the HELO$\times$1 extended basis, and less than 1~meV for the other 
basis sets. Energy differences are converged to an accuracy that is one order of magnitude higher than 
required in the test calculations. Figure~\ref{dosBandsCeFigure} shows
the density of states (DOS) and the band structure along the high-symmetry lines K$-\Gamma-$L 
together with the energy parameters obtained from the conventional LAPW basis. 
The energy parameters all lie within their associated bands and within a 
distance of only a few eV to the occupied valence states, which cover an energy interval of
$3~\mathrm{eV}$ below the Fermi level.

The second material, KCl in the rock-salt structure, features a large KS band gap of 
about $5~\mathrm{eV}$. To describe this gap accurately, electronic states over a broad energy range have to 
be represented accurately. Figure~\ref{dosEParamKClFigure} shows the density of states together with the energy parameters. The band 
structure is investigated in detail in Sec.~\ref{unoccupiedStates} and is shown in 
Fig.~\ref{bandStructureKClFigure}. We remark that the Cl $3p$ states give rise to
small $d$ and $f$ contributions to the DOS in K, which are energetically far away from the $E_{3d,\mathrm{K}}$ 
and $E_{4f,\mathrm{K}}$ energy parameter, respectively (cf.\ Table~\ref{calculationParametersTable}).

\begin{table}
\caption{Calculational parameters used unless stated otherwise. For each material, the table lists the crystal structure, the applied experimental lattice constant, the MT radii ($R_\mathrm{MT}$), the reciprocal plane wave cutoff as 
$G_{\mathrm{max}}R_{\mathrm{MT}}$, the number of LAPW basis functions (averaged over the $\vec{k}$ points), the angular momentum cutoff $l_\mathrm{max}$, the number of $\vec{k}$ points in the IBZ, the semicore LOs, and the obtained atomic energy 
parameters relative to the Fermi level $E_{l}-E_\mathrm{F}$ for the conventional LAPW 
basis. Energy parameters for $l>3$ are set to $E_{l=3}$.}
\begin{tabular*}{\columnwidth}{l c c c c}
\hline
\hline
parameter & Ce & KCl & Ar & V \\
          &    & (K, Cl)       \\
\hline
crystal structure             & fcc        & rock-salt    & fcc        & bcc \\
latt. const.~($a_0$)          & $9.05$ & $11.89$ & $9.93$ & $5.73$ \\
\hline
$R_\mathrm{MT}~(a_0)$ & $3.14$ & $2.8$, $2.8$ & $3.15$ & $2.41$ \\
$G_{\mathrm{max}}R_{\mathrm{MT}}$ & 13 & 13 & 13 & 10.845 \\
LAPWs / atom & 222 & 355 & 295 & 145 \\
$l_\mathrm{max}$ & 12 & 12 & 10 & 10 \\
$\vec{k}$ points in IBZ & 182 & 60 & 60 & 190 \\
semicore LOs & $5s$,$5p$ & $3s$,$3p$ (K) \\
\hline
\multicolumn{5}{c}{atomic energy parameters~(eV)} \\
$E_{l=0}-E_\mathrm{F}$   & $-$1.18 & 6.14, 11.95 & $-$14.43 & $-$3.34\\
$E_{l=1}-E_\mathrm{F}$   & 1.81 & 8.97, $-$0.39 & $-$0.49 & $-$0.22\\
$E_{l=2}-E_\mathrm{F}$   & 0.88 & 9.27, 10.88 & 14.78 & 0.05\\
$E_{l\ge3}-E_\mathrm{F}$ & 0.73 & 15.63, 16.08 & 20.56 & 2.05\\
\hline
\end{tabular*}
\label{calculationParametersTable}
\end{table}

\begin{figure}
\centering
\includegraphics*[width=0.8\columnwidth]{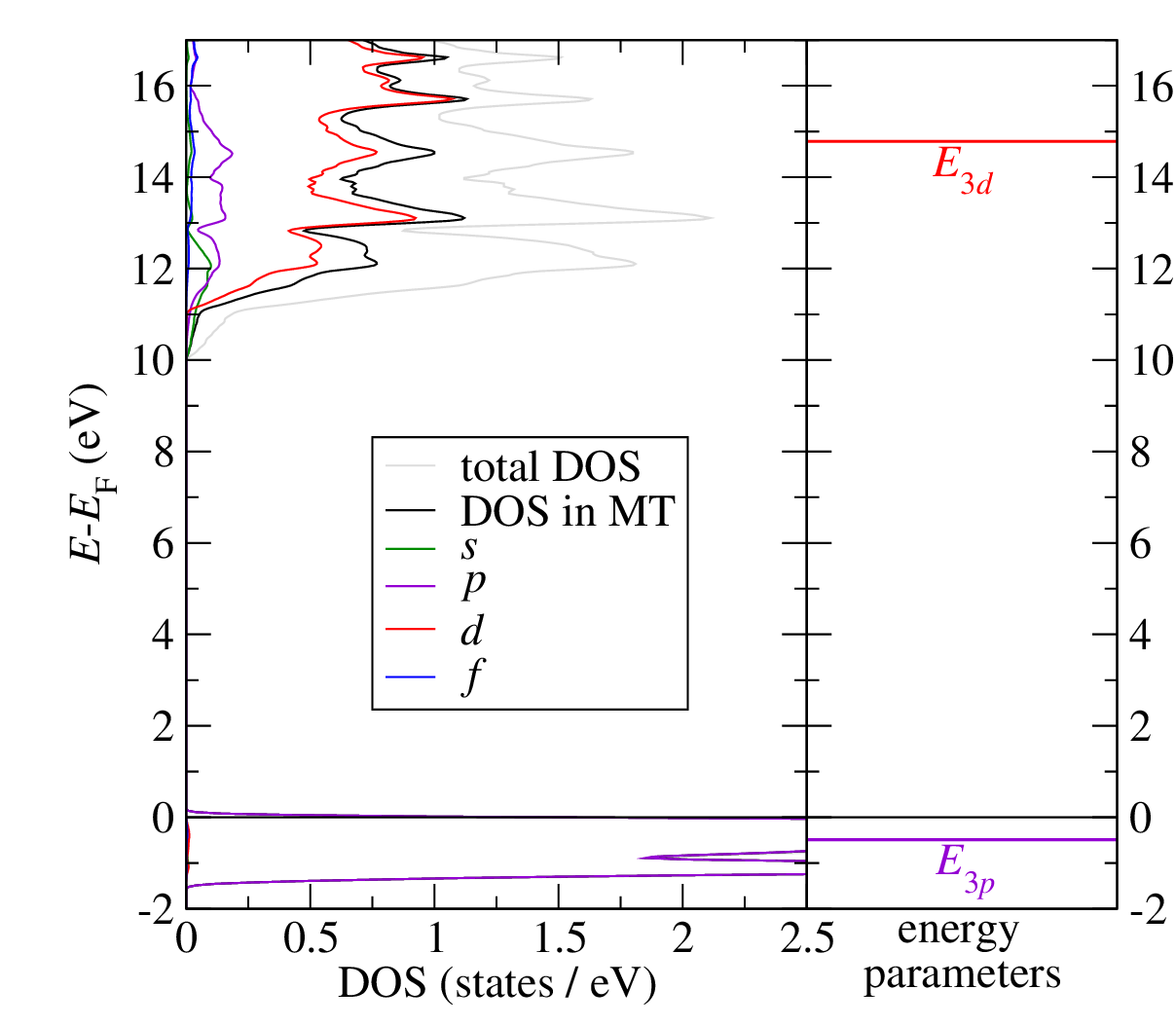}
\caption{Density of States and atomic energy parameters for Ar.}
\label{dosEParamArFigure}
\end{figure}

Fcc Argon is the third material that we investigate. It also features a large band gap and enables the usage of 
a large MT radius. Density of states and energy parameters for this material are shown in 
Fig.~\ref{dosEParamArFigure}. Similarly to KCl, we discuss the band structure of Ar in detail in 
Sec.~\ref{unoccupiedStates} and show it in Fig.~\ref{bandStructureArFigure}.

To demonstrate the description of semicore states by the different basis sets we also perform 
calculations on bcc vanadium.

\section{Linearization error}
\label{theLinearizationError}

In this section we quantify the linearization error for the different basis sets. 
First, we define a basis representation error that quantifies the linearization 
error and examine its dependence on the energy parameters and MT sphere radii. 
Up to that point, we only consider non-selfconsistent calculations, where
the ability of the basis to represent the exact single-particle wavefunctions in the
spheres to a given effective potential is assessed.
Then, we turn to self-consistent calculations. In particular, we examine how
total energies, lattice constants, and KS band gaps depend on the energy parameters and the
MT sphere radii. We will see that the LO extension reduces these undesirable
dependences to a great extent.

\subsection{MT basis representation}
\label{representationError}

In order to assess the quality of the basis representation in the MT spheres,
we define the representation error
\begin{equation}
\Delta_{l}(E) = \Vert u_l - \tilde{u}_l\Vert= \left( \int r^2 \left[u_l(r,E)-\tilde{u}_l(r,E)\right]^2 dr \right)^\frac{1}{2}\,,
\label{deflinerror}
\end{equation}
where $u_l(r,E)$ is a normalized solution of Eq.~\eqref{udiff} for a given energy $E$ and 
$\tilde{u}_l(r,E)$ is the \emph{best} representation of $u_l(r,E)$ in terms of the radial 
functions of the given basis in the MT sphere, i.e, $\tilde{u}_l(r,E)-u_l(r,E)$ is
orthogonal to the basis. 
%Thus, r
The error $\Delta_{l}(E)$ becomes 
zero if $u_l(r,E)$ can be represented pointwise by the basis. This is the case
if $E$ is identical to an energy parameter. On the other hand, 
$\Delta_{l} = 1$ if $u_l(r,E)$ is orthogonal to the function space spanned 
by the available radial functions. Note that the representation error defined above only covers that part of the wave functions 
that is inside the MT sphere. It does not include the additional boundary 
conditions imposed by the matching of the radial functions to the plane
waves at the MT boundary. This matching
reduces the flexibility of the basis such that the representation error of ${u}_l$ represents a
lower bound for the error with respect to the entire LAPW basis.

\begin{figure}
\centering
\includegraphics*[width=\columnwidth]{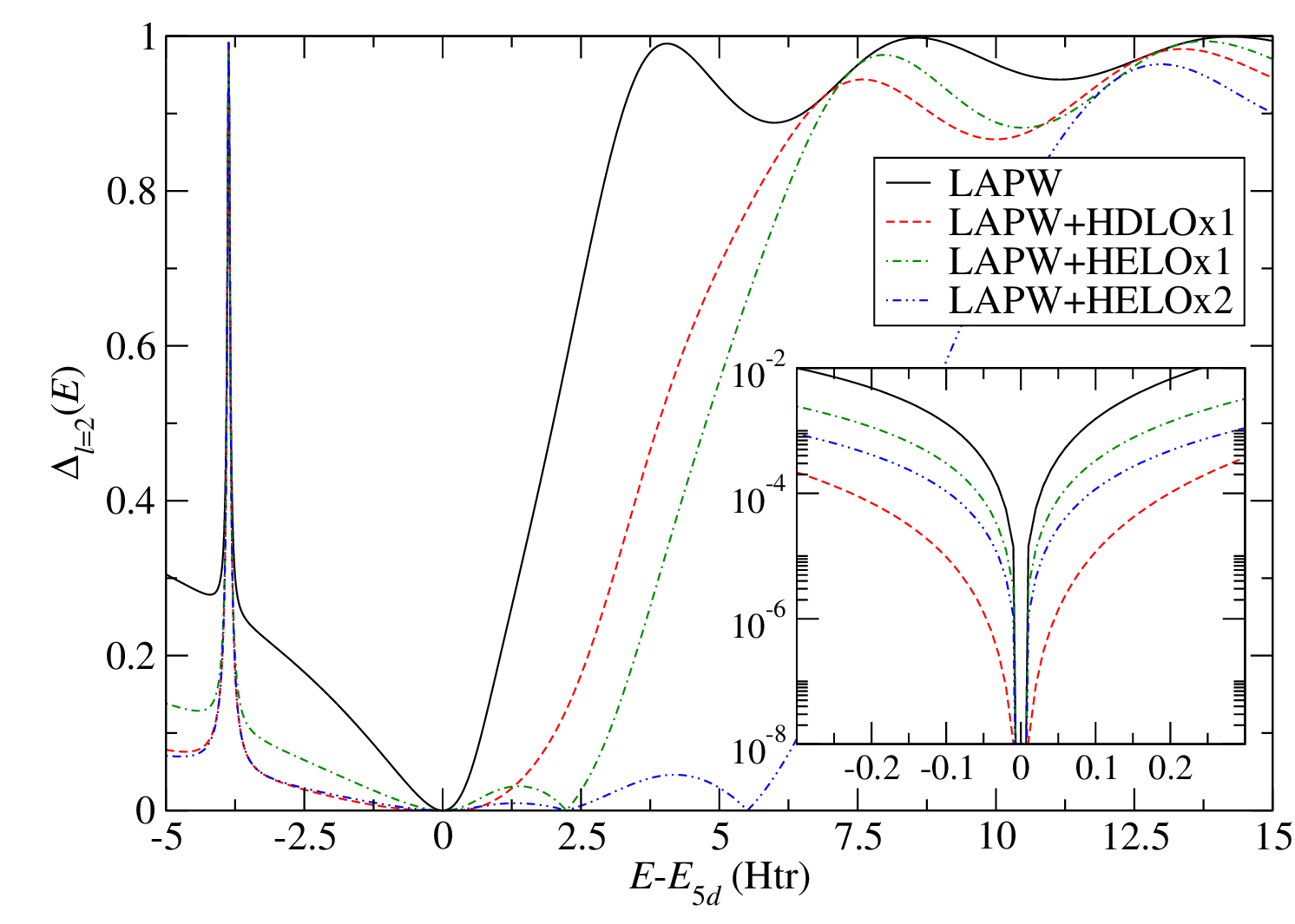}
\caption{Representation error in the $d$ channel of the MT sphere in fcc Ce. 
There is a localized $d$ core state at $E-E_{5d}=-3.87
\thinspace\text{Htr}$, which is orthogonal to the space spanned by the basis
sets. The inset shows a detailed view around the energy parameter 
in a logarithmic scale.}
\label{energyDependenceCeFigure}
\end{figure}

In the following, we investigate the energy dependence of the representation 
error $\Delta_{l}(E)$ for the conventional LAPW basis and its 
HDLO and HELO extensions. Based on the effective potential from a
self-consistent DFT calculation, Figure~\ref{energyDependenceCeFigure} shows the 
error for $l=2$ in the vicinity of the energy parameter for 
the $5d$ states of bulk fcc Ce.

For an accurate representation of the electron density, which is the central 
quantity of DFT, the valence states must be described accurately. Therefore,
the energy parameters are chosen in the valence band. 
The inset of Fig.~\ref{energyDependenceCeFigure}
shows an energy interval of about $\pm0.3~\mathrm{Htr} 
\approx \pm8~\mathrm{eV}$ around the Ce $5d$ energy parameter.
In this energy range the linearization error of 
the conventional LAPW basis for the $d$ channel of Ce amounts 
to maximally $1.5 \cdot 10^{-2}$. This error can be further 
reduced by adding LOs. While the HELO$\times$1 and HELO$\times$2 basis sets reduce the
maximal error to $3.3 \cdot 10^{-3}$ and $1.1 \cdot 10^{-3}$, respectively, the
LAPW+HDLO$\times$1 exhibits a maximal error of only $3.8\cdot 10^{-4}$, a factor of
40 smaller than in the case of the conventional LAPW basis.
We note that in the immediate vicinity of the energy parameter,
$\Delta_{l}(E)$ scales as $|E-E_l|^2$ for the conventional LAPW basis 
and as $|E-E_l|^3$ for the HDLO$\times$1 basis.

For energetically high-lying states the conventional LAPW basis quickly becomes inadequate, 
while LOs, especially the HELOs (cp.~Fig.~\ref{energyDependenceCeFigure}), can improve the basis
substantially in a systematic and controllable manner. Methods that rely on
the empty states such as the $GW$ approximation~\cite{PhysRev.139.A796} or RPA correlation
functionals~\cite{RPA-Review-1,RPA-Review-2} thus require an LAPW basis that is thoroughly converged with
respect to LOs.

\begin{figure}
\centering
\includegraphics*[width=\columnwidth]{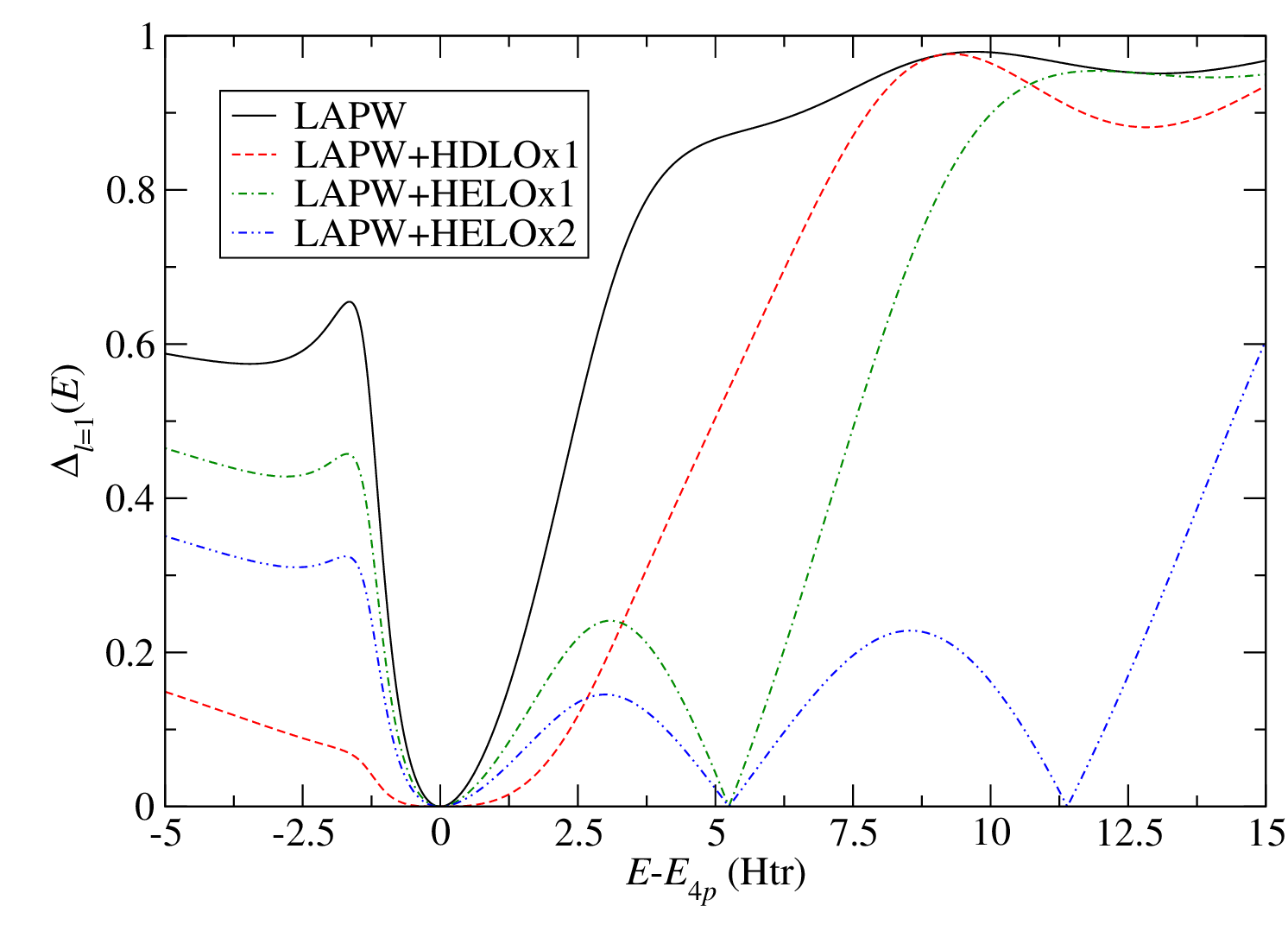}
\caption{Same as Fig.~\ref{energyDependenceCeFigure} for the $p$ channel in bcc V. 
The semicore state at $-1.7\thinspace\text{Htr}$ is not completely confined 
in the MT sphere. Therefore, it is not orthogonal to the space spanned by
the basis sets and appears as shallow peaks with amplitudes well below 1.}
\label{energyDependenceVFigure}
\end{figure}

For all basis sets we observe a sharp peak at $-3.87~\mathrm{Htr}$ ($=-105~\mathrm{eV}$) where
$\Delta_l(E)$ becomes 1, which signals a state that is orthogonal to the
basis functions and therefore cannot be described by the basis. 
In practice, this is no problem as this is the $4d$ core state of Ce which is treated separately from the valence states. 
As it is completely confined to the 
MT sphere, it lies outside the space spanned by the basis sets. 
On the other hand, the wave function of a semicore state extends
significantly over the MT sphere. In particular, it does not vanish
at the boundary and will therefore have a finite overlap with the valence
basis. Figure \ref{energyDependenceVFigure} shows such a case in the $p$
channel of bcc vanadium. The representation error features a shallow peak
(or merely a shoulder) at $-1.7~\mathrm{Htr}$ ($=-46~\mathrm{eV}$), which corresponds to the $3p$ state.
This impedes a treatment of the $3p$ state separate from the valence states,
and we must employ LOs of $l=1$ at the appropriate energy, especially if 
the LAPW+HDLO$\times$1 basis is applied. We note that although the semicore state could 
also be described by systematically including LOs with higher and higher orders of derivatives, 
the more traditional way~\cite{PhysRevB.43.6388} of employing a single LO, constructed from the solution to Eq.~(4) 
with $E$ being the semicore energy, is much more efficient since the semicore states form 
very flat bands with hardly any dispersion and should therefore be readily representable 
by a single LO. It is evident that the rest of the diagram looks rather similar to the one in 
Fig.~\ref{energyDependenceCeFigure}(a), which demonstrates that the
qualitative behavior of the representation error is essentially independent of the
material and the angular momentum $l$.

While the energy parameters can be adapted to the material in each iteration of the
self-consistent-field run, the MT radius must be
chosen before starting the calculation and then stays at this fixed value. 
One condition for the choice of the radius is
that the MT spheres must not overlap. On the other hand, they
should be chosen as large as possible in order to keep the reciprocal cutoff
radius $G_\mathrm{max}$ small. (The larger the spheres, the smaller the IR,
and the less plane waves are necessary to represent the wave functions there.)
Finally, if one considers more than one chemical element, the sizes of the MT spheres relative to each other
can be chosen according to tabulated atomic radii. This shows that
there is no optimal choice of the radii. Ideally, the calculated results
should be independent of the choice of the MT radii.

\begin{figure}
\centering
\includegraphics*[width=\columnwidth]{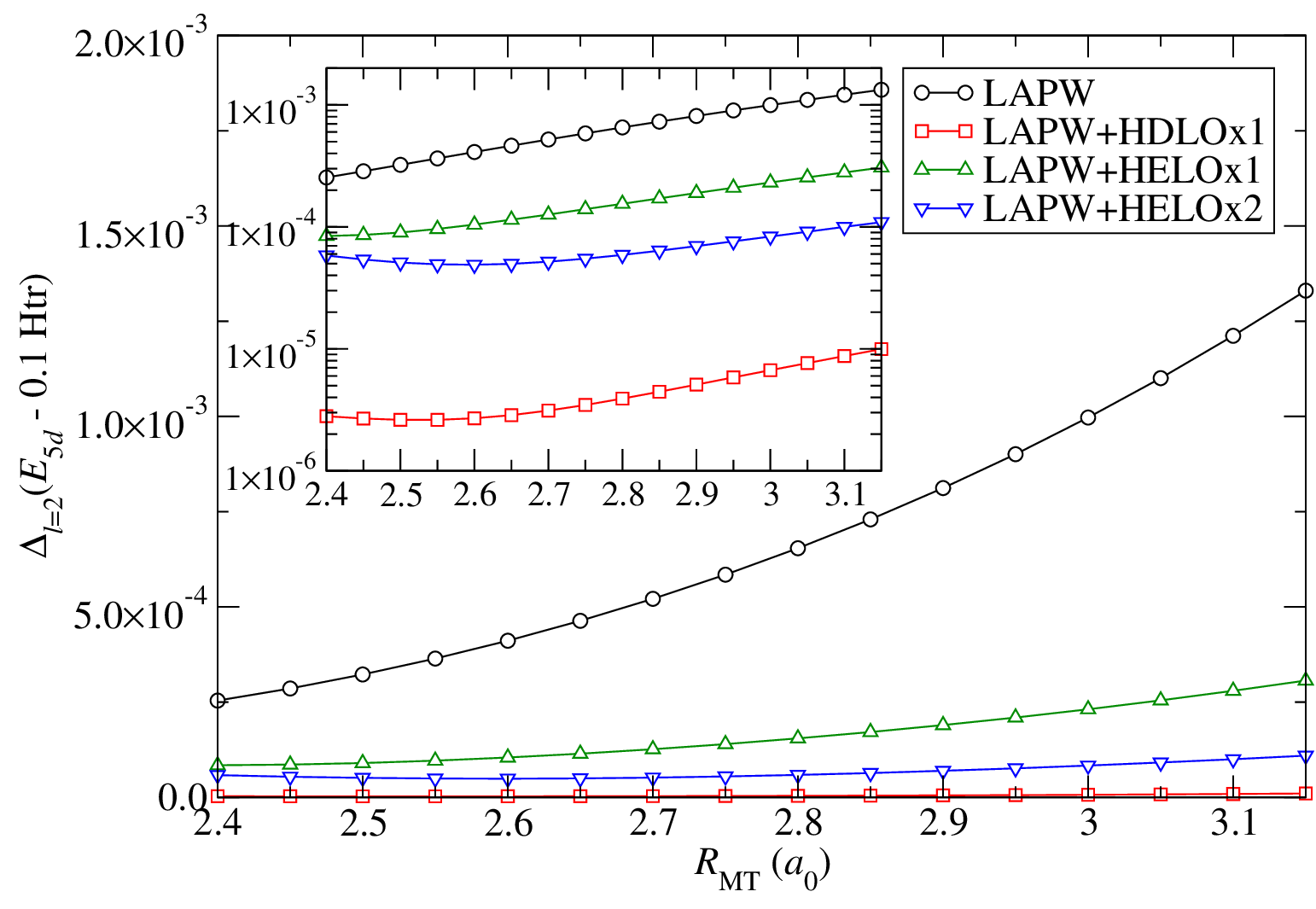}
\caption{Representation error at the fixed energy $E_{5d}-0.1~\mathrm{Htr}$ 
as a function of the MT radius in fcc Ce. The LAPW+HDLO$\times$1 basis
suppresses the representation error most efficiently. The error is
below $10^{-5}$. Also compare the logarithmic plot in the inset.}
\label{rmtReductionLinErrorFigure}
\end{figure}

In Fig.~\ref{rmtReductionLinErrorFigure} we show the representation error 
for $5d$ states in Ce as a function of the MT radius for the different basis sets and 
at a fixed energy $E$ of $0.1~\mathrm{Htr}$ below the $5d$ energy parameter. 
In the calculations we increased the cutoff $G_{\mathrm{max}}$
while reducing the MT radius to keep the product $G_{\mathrm{max}}R_{\mathrm{MT}}$ 
fixed to $13$. The effective potential was determined from self-consistent
calculations for each MT radius. All other numerical parameters, as well as the 
method to choose the energy parameters, are identical to the previous Ce calculation.

We observe that, independently of the basis chosen, the representation error 
becomes smaller as the MT radius is reduced. This is related to the growing
independence of $u_l(r,E)$ on the energy parameter $E$ with decreasing radius observed in 
Fig.~\ref{logDerivCeFigure}(b) and is most pronounced for the 
conventional LAPW basis, which has no additional LOs apart from those for the
semicore states. From $3.15~a_0$ to $2.4~a_0$ the error decreases from
$1.3\cdot 10^{-3}$ to $2.5\cdot 10^{-4}$.
In the case of the LAPW+HELO and LAPW+HDLO basis sets, 
the dependence on the MT radius is strongly reduced. In particular, the 
LAPW+HDLO$\times$1 basis exhibits a very small representation error for this
wide range of MT radii, nearly a factor 100 smaller than in
the conventional LAPW basis. Such a stability with respect to non-convergence 
parameters like MT radii is clearly desired.

\subsection{Total energies and lattice constants}
\label{calculationDependencesSection}

Now we move to self-consistent calculations and address the questions whether 
the observations of the previous section
translate to physical properties, such as the total energy $E_\textrm{total}$ or
the equilibrium lattice constant $a$ after self-consistency in the electron density is achieved.
Thus, we write the total energy $E_\textrm{total}(\tilde{a}\, | \,\{E_l\},\{R_\textrm{MT}\})$ in a form where besides 
its dependence on the lattice constant $\tilde{a}$, the number of Bloch vectors, and the size 
of the basis set, which both have been chosen sufficiently large that the total energy can be considered converged with
respect to these parameters, its dependence 
on the set of energy parameters and MT radii becomes explicit. It is clear that the equilibrium lattice constant 
$a$ is obtained by minimization of the total energy with respect to variations of the lattice constant.
The total energy $E_\textrm{total}$ itself is dependent on the method-inherent parameters.
As in the previous section, we focus on the stability of the results with
respect to variations of the energy parameters and the muffin-tin radii.

In several respects, self-consistent calculations are more demanding on the performance of basis sets than what we have considered
so far. Firstly, we now account for the full
nonspherical potential in the MT spheres, while we have restricted ourselves
to the spherical potential before. Second, the IR is now taken into
account as well, and the wave functions must be described accurately over the
complete space. Third, the wave functions must also be described accurately over a sufficiently 
wide energy range to yield a precise valence electron density in each
self-consistent-field iteration. 

\begin{figure}
\centering
\includegraphics*[width=\columnwidth]{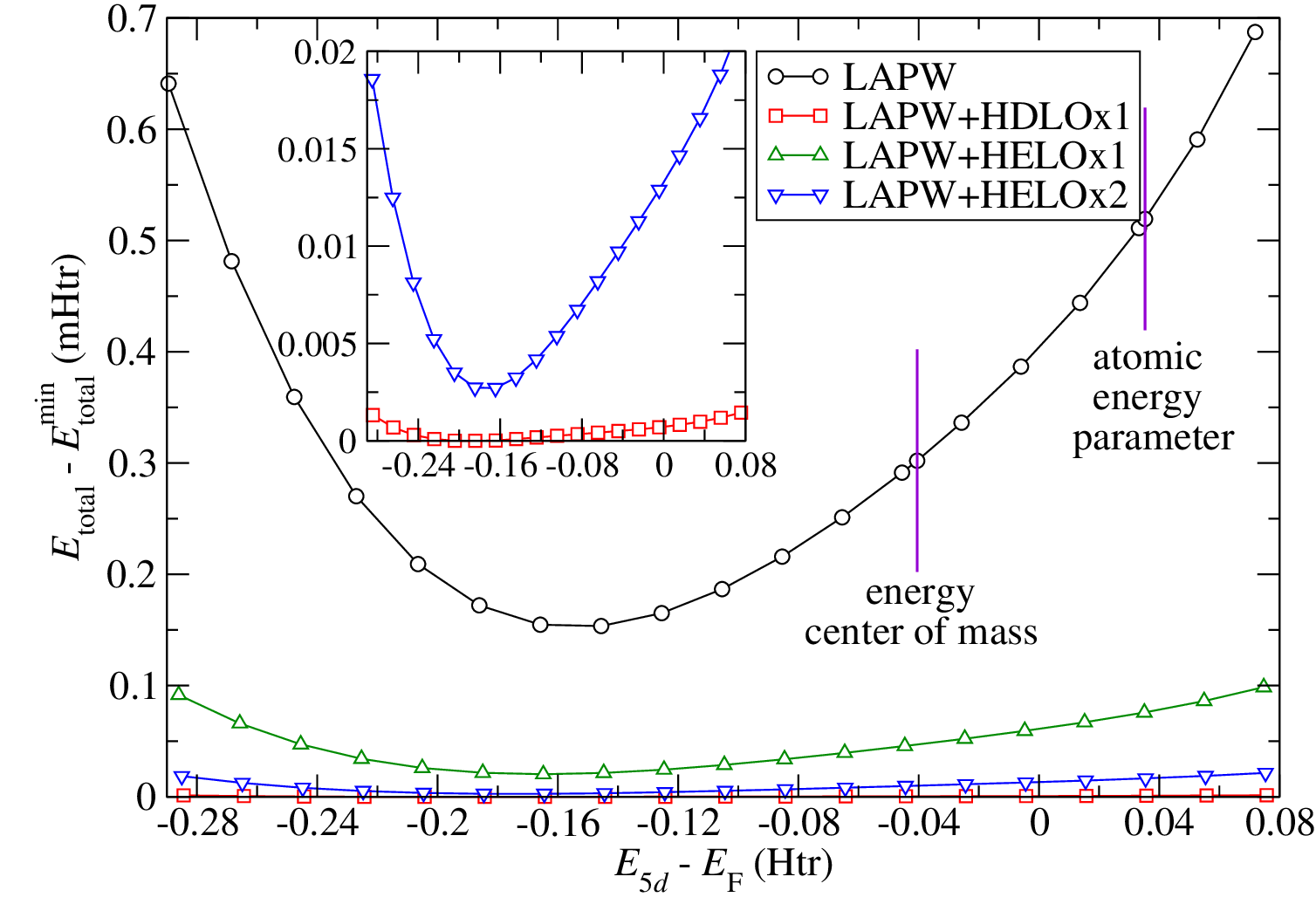}
\caption{Total energy of fcc Ce for the different basis sets from
self-consistent-field calculations as a function of the
energy parameter $E_{l=2}$ relative to the Fermi energy.
The total energy is plotted relative to the minimal energy obtained with the basis set including HDLOs. 
The inset shows a detailed view, revealing that the LAPW+HDLO$\times$1 basis
exhibits the smallest error by far. We also mark the energetic positions
of the energy parameter as determined from two common approaches, which are described 
in the text. In these calculations HDLOs and HELOs are only added in the $d$ channel.}
\label{energyParameterChoiceFigure}
\end{figure}

In Fig.~\ref{energyParameterChoiceFigure} we show the dependence of the
calculated total energy of fcc Ce on the energy parameter $E_{l=2}$
for different basis sets, where, as an exception, the LAPW basis for the valence electrons is only extended
by LOs of $d$ character. While keeping all other energy parameters fixed
(to values taken from a previous self-consistent-field calculation),
we vary $E_{l=2}$ in the range from $-0.29$ to $0.08~\mathrm{Htr}$ ($\approx-7.9$ to $2.2~\mathrm{eV}$) 
relative to the Fermi level. As seen in the figure, the total energy 
calculated with the conventional LAPW basis set depends 
significantly on the choice of the energy parameter. The marked energy
parameters determined from the two automatic approaches yield total
energies that deviate substantially from each other and from the minimal total energy obtained with 
$E_{l=2}-E_\textrm{F}=-0.16~\mathrm{Htr}$ which lies about $1.3~\mathrm{eV}$ below the lower valence band edge.

When extending the LAPW basis with LOs, we observe that the total energy
becomes lower (due to the increase of the variational freedom enabled by
the larger basis sets), but, more important, its dependence on $E_{l=2}$ is
strongly suppressed, most strongly again for the LAPW+HDLO$\times$1 set,
which exhibits a variation that is two orders of magnitude smaller than for
the conventional LAPW basis and, thus, negligible for all practical purposes.
Although the HELO$\times$2 extension contains one set of LOs more than
HDLO$\times$1, the corresponding total energy exhibits a larger variation
albeit still a very small one when
compared to the conventional set. Of course, the calculations with the HELOs still
depend on $E^{\mathrm{LO}}_{l=2}$, i.e, the energy parameter of the
HELOs. As already explained above, we employ for the HELOs energy 
parameters determined from Eq.~\eqref{logder}. The HDLO accuracy 
could be realized with normal LOs if their energy parameter 
$E_{l=2}^\mathrm{LO}$ was chosen close to $E_{l=2}$. In fact, in the limit 
$E_{l=2}^{\mathrm{LO}} \rightarrow E_{l=2}$ the two approaches are theoretically identical. However, if the 
distance between the two energy parameters gets too small, the near linear dependence of the 
basis might easily cause numerical problems, which is avoided by using HDLOs.

\begin{figure}
\centering
\includegraphics*[width=\columnwidth]{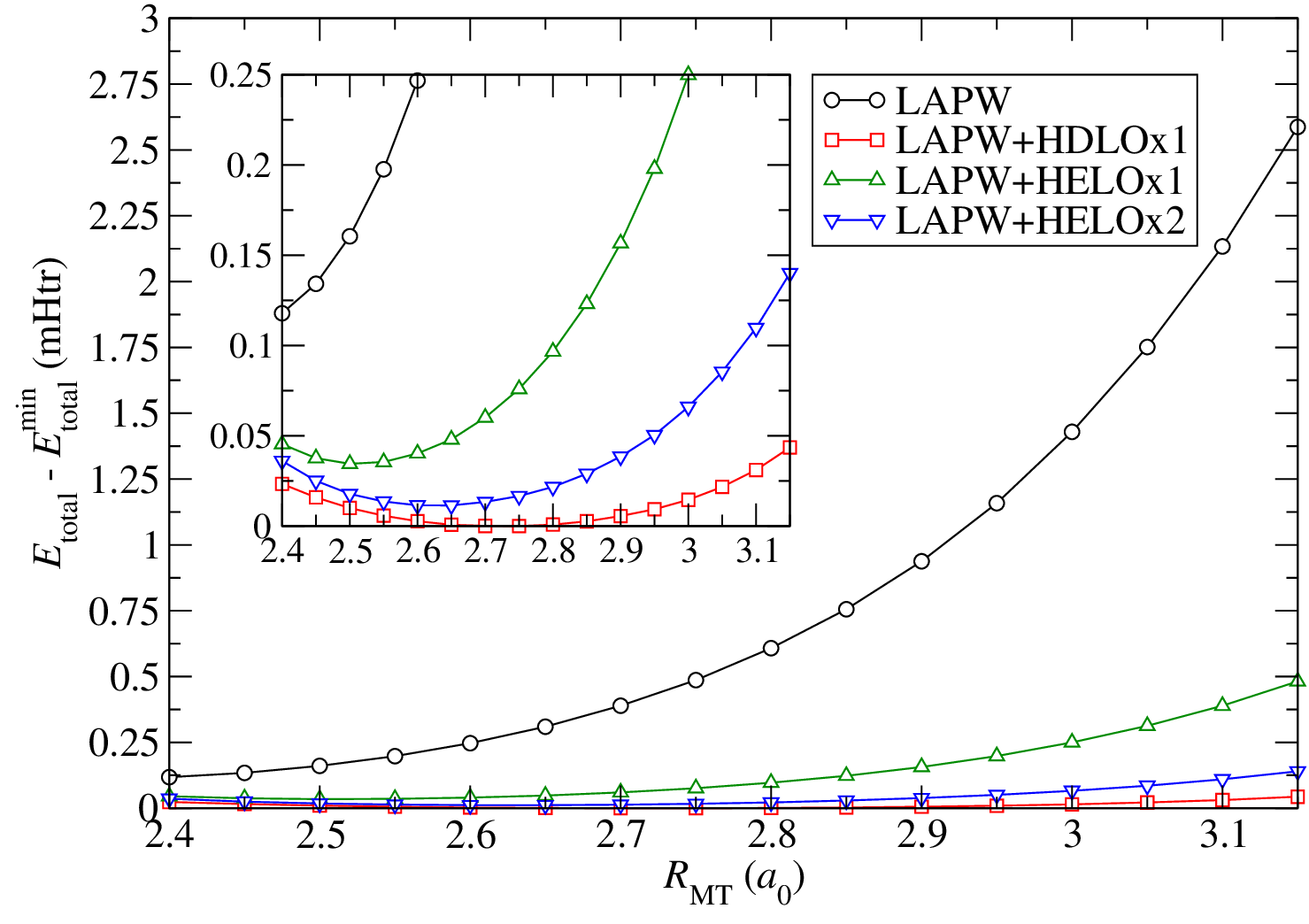}
\caption{Ground-state total energy of fcc Ce from
self-consistent-field calculations as a function of the
MT radius for different basis sets. 
HDLOs and HELOs are added for $l=0,1,2,3$.
The inset shows again that the LAPW+HDLO$\times$1 basis
exhibits the smallest error.}
\label{totEnergyVsRmtCeFigure}
\end{figure}

\begin{figure}
\centering
\includegraphics*[width=\columnwidth]{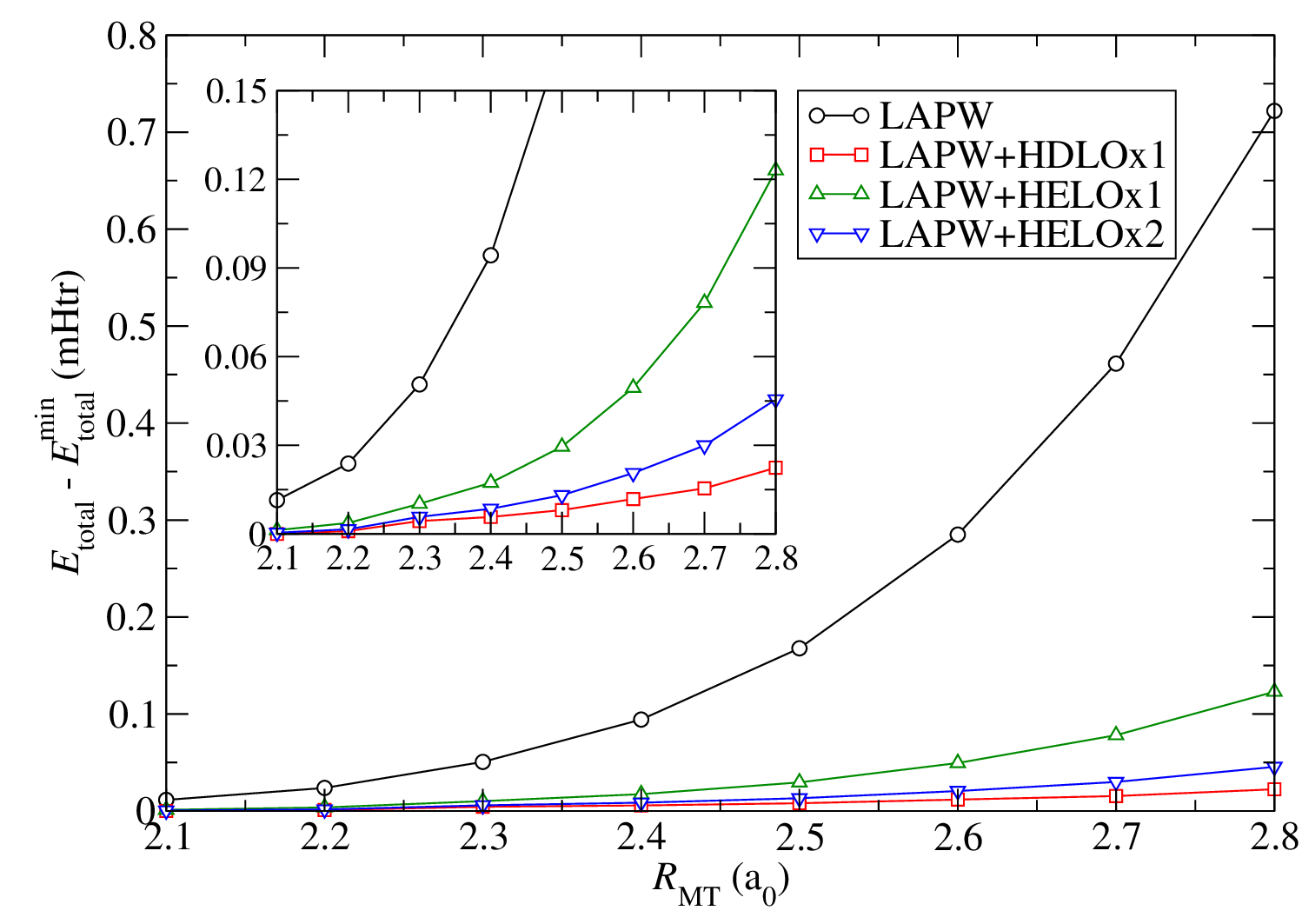}
\caption{Same as Fig.~\ref{totEnergyVsRmtCeFigure} for KCl.}
\label{totEnergyVsRmtKClFigure}
\end{figure}

Besides its dependence on the energy parameters, in Fig.~\ref{rmtReductionLinErrorFigure} we already showed that the linearization 
error also strongly depends on the MT radii. The error decreases for smaller
radii, but then the IR becomes larger,
which entails a larger reciprocal cutoff radius giving rise to an increase
of the basis set size and thus higher computational demands. As we have seen, a more effective means to
reduce or even eliminate the linearization error is the introduction of LOs, in particular
the HDLOs. 

We observe the same behavior for the total energy
and the equilibrium lattice constant, for which the total energy assumes a
minimum. The latter is calculated from a series of
total-energy calculations for different lattice constants together with a
Murnaghan fit \cite{pnas01666-0028}. We show results for Ce and KCl. 
To avoid overcompleteness of the basis at small MT radii, we neither include 
HDLOs nor HELOs in the $s$ and $p$ channels of K, where we already 
have LOs to describe the semicore states.

In Figs.~\ref{totEnergyVsRmtCeFigure} and \ref{totEnergyVsRmtKClFigure} 
the total energy is shown as a function of the MT radius. In fact, the diagrams look
qualitatively very similar to the MT representation error shown in
Figs.~\ref{energyDependenceCeFigure} and \ref{rmtReductionLinErrorFigure}, 
respectively. The variation in the total energy is between one
and two orders of magnitude smaller with the HDLO extension than without for
both compounds.

\begin{figure}
\centering
\includegraphics*[width=\columnwidth]{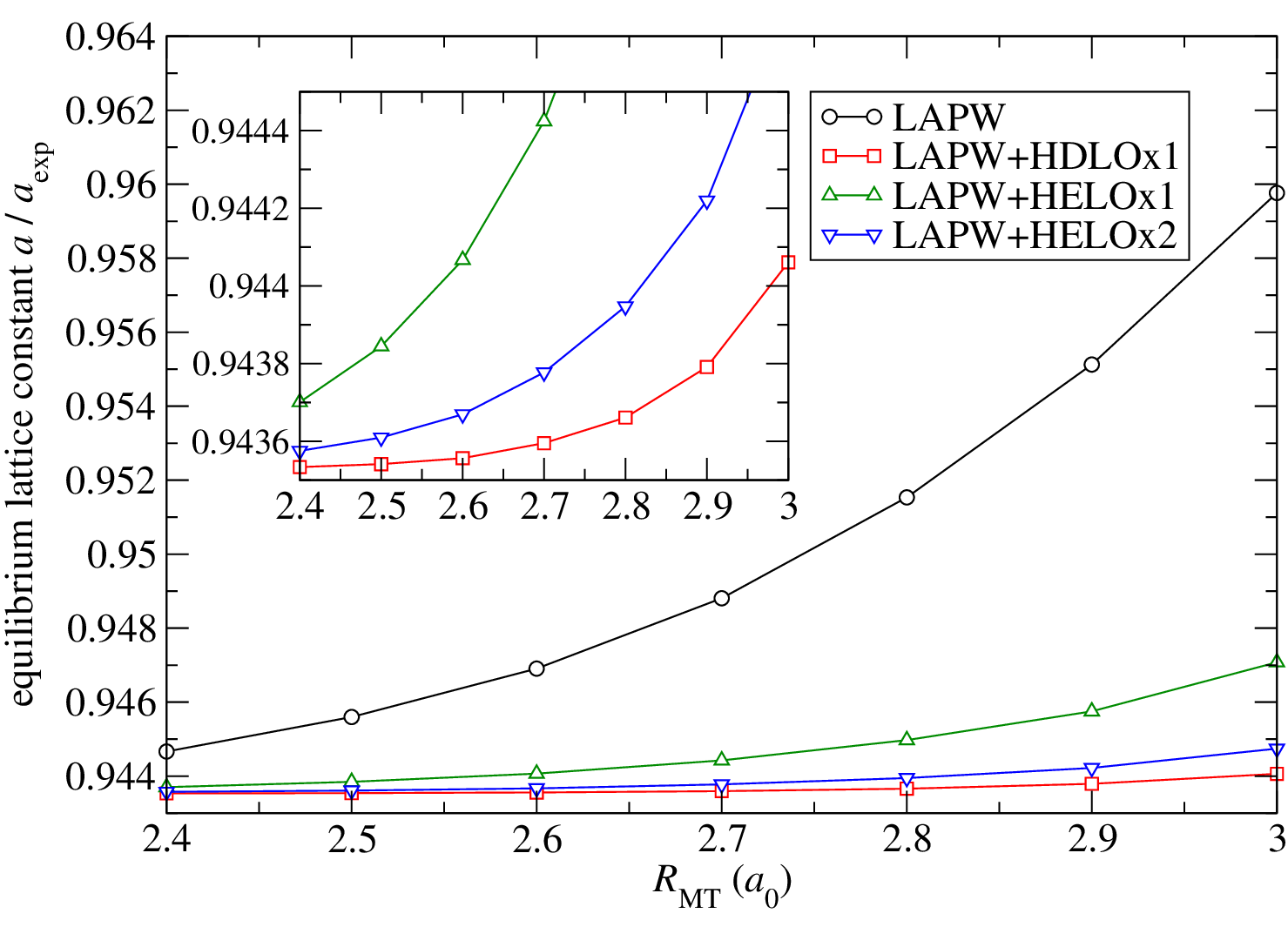}
\caption{Equilibrium lattice constant as a function of the MT radius for 
fcc Ce determined from Murnaghan fits to the total energies
for 15 lattice constants in the range $0.938 a_\text{exp}$ to $0.966
a_\text{exp}$.}
\label{lattConstVsRmtCeFigure}
\end{figure}

\begin{figure}
\includegraphics*[width=\columnwidth]{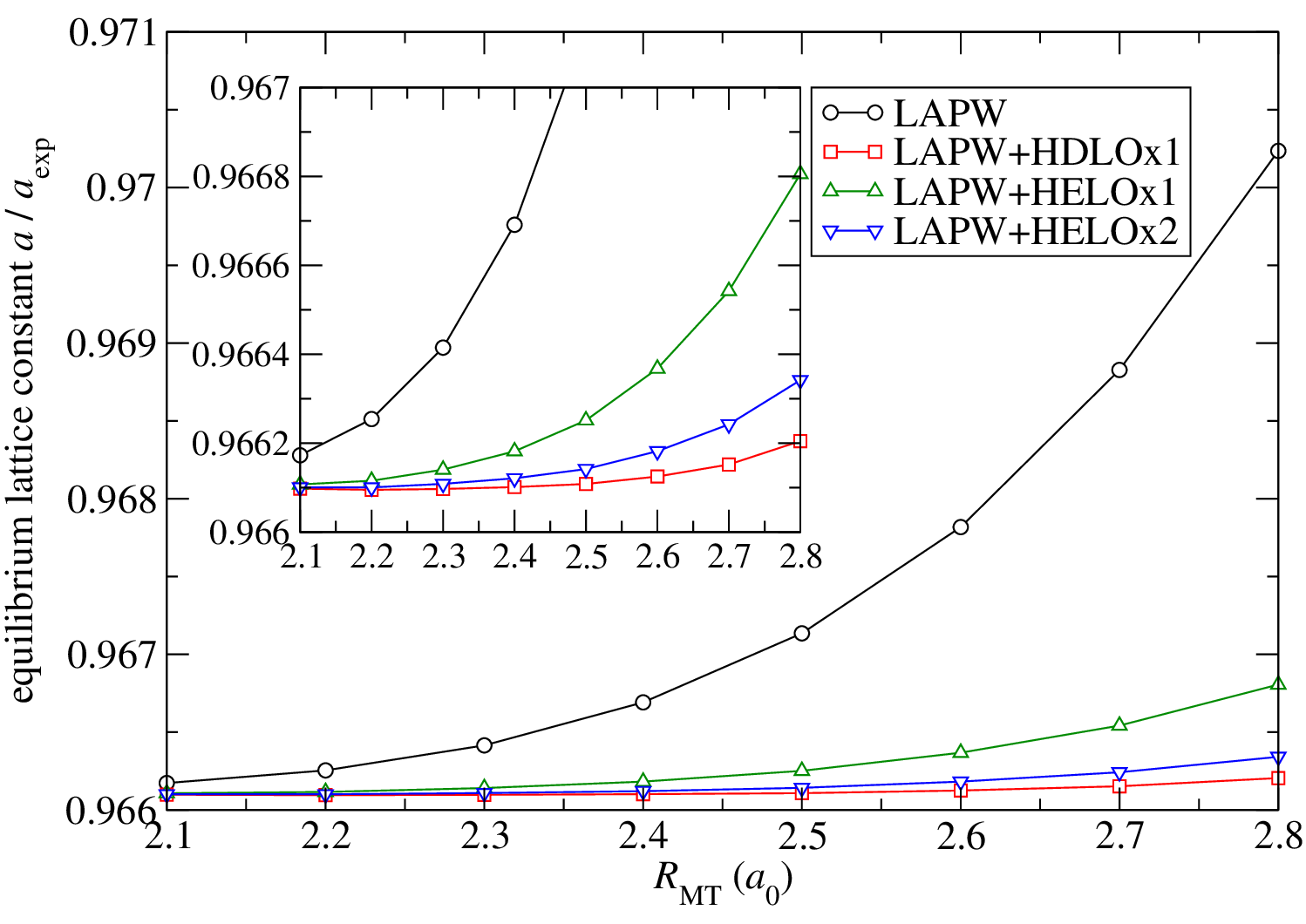}
\caption{Same as Fig.~\ref{lattConstVsRmtCeFigure} for KCl. Nine
total-energy calculations for lattice constants in the range 
$0.958 a_\text{exp}$ to $0.974 a_\text{exp}$ were employed for the fit.}
\label{lattConstVsRmtKClFigure}
\end{figure}

Even without the LOs, the total energy is accurate up to few mHtr.
However, the equilibrium lattice constant determined from the
minimum of the total energy depends on small energy
differences. It is thus particularly susceptible to inaccuracies in the total energies.
As shown in Fig.~\ref{lattConstVsRmtCeFigure} and
\ref{lattConstVsRmtKClFigure}, the variation of the lattice constant $a$ is on the order of
1 percent for Ce ($a_\mathrm{exp}=9.05~a_0$, $a_\mathrm{LAPW}= 8.55$ to $8.69~a_0$) and KCl ($a_\mathrm{exp}=11.89~a_0$, $a_\mathrm{LAPW}=11.49$ to $11.54~a_0$) and thus in the same ballpark as the supposed 
accuracy of the LDA functional. After addition of one set of HDLOs the
variation of $a$ is strongly suppressed to less than 0.1\% (Ce: $a_\mathrm{LAPW+HDLO\times1}= 8.539$ to $8.544~a_0$, KCl: $a_\mathrm{LAPW+HDLO\times1}=11.487$ to $11.488~a_0$).

We note that each additional set of LOs (for $l=0, {\ldots}, 3$) increases the basis-set
size by only 16 functions per atom. Thus, adding LOs is 
computationally much more efficient than reducing the MT radius, which would 
make many more augmented plane waves necessary to enable
an adequate description of the wave functions in the interstitial region. 

\subsection{KS band gap}
\label{unoccupiedStates}

The band gap is a fundamental quantity of
semiconductors and insulators and is given by the energy difference of the lowest
unoccupied and the highest occupied state. Its calculated value also depends on the
ability of the basis set to describe both, the valence and the conduction bands with sufficient accuracy. Since the energy parameters
are located typically in the energy range of the valence bands (see Sec.~\ref{calculationDependencesSection}), 
one might expect deviations of the conduction states from the true KS eigenstates 
having an adverse effect on the accuracy of the band-gap value.
We demonstrate for KCl and Ar that the inaccuracies of the KS gap due to the
linearization error can indeed be sizable unless LOs are used to
ensure a proper description in the MT spheres. With the LAPW basis set,
the KS band gap is calculated to be larger than with the extended basis sets,
leading to an underestimation of the KS band gap error when compared to 
experimental values.

\begin{figure}
\centering
\includegraphics*[clip,width=\columnwidth]{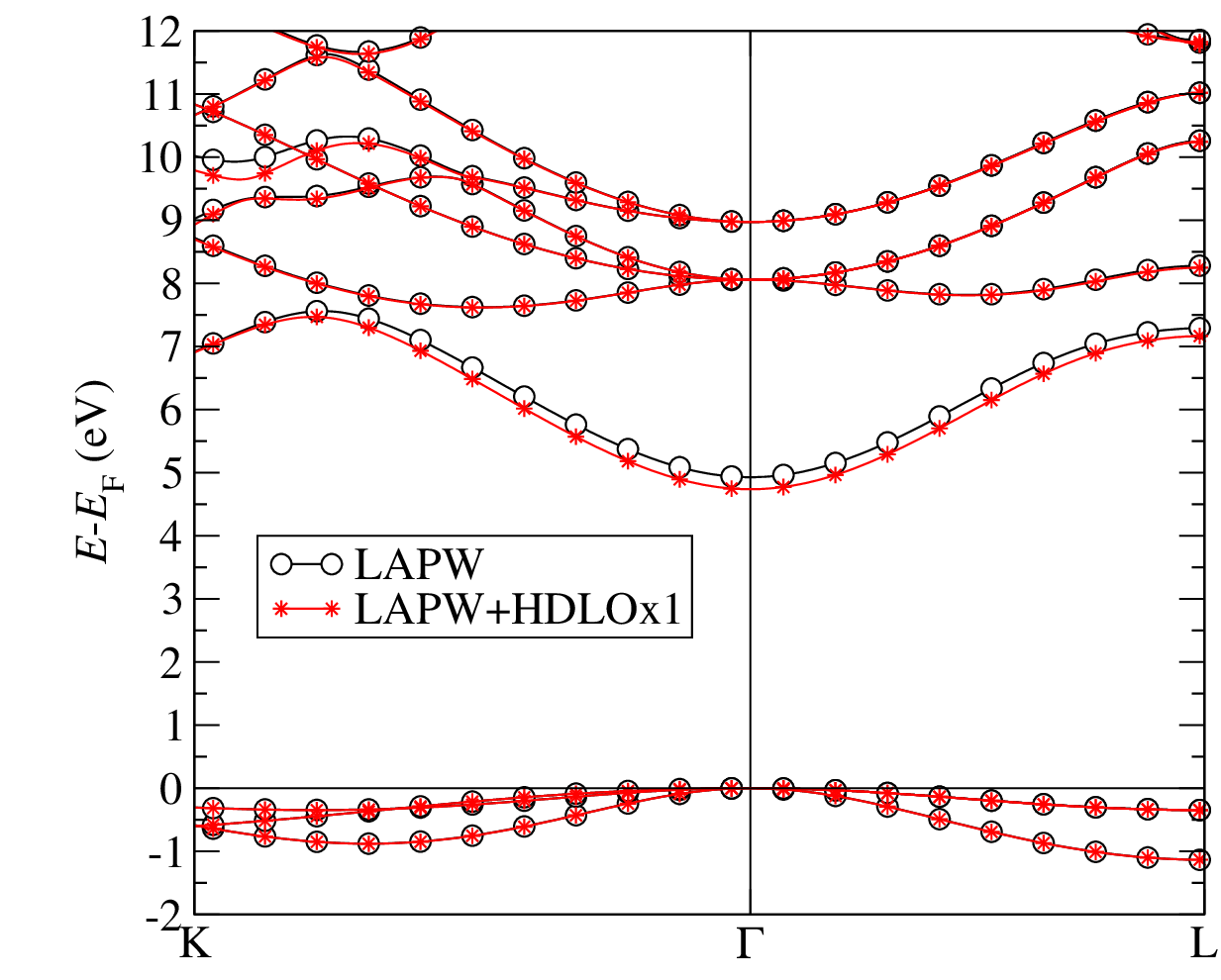}
\caption{Band structure of KCl calculated with the conventional LAPW
and the LAPW+HDLO$\times$1 basis set. On the given energy scale the band structures calculated with the HELO
extensions are indistinguishable from the one with HDLOs.}
\label{bandStructureKClFigure}
\end{figure}

\begin{figure}
\centering
\includegraphics*[clip,width=\columnwidth]{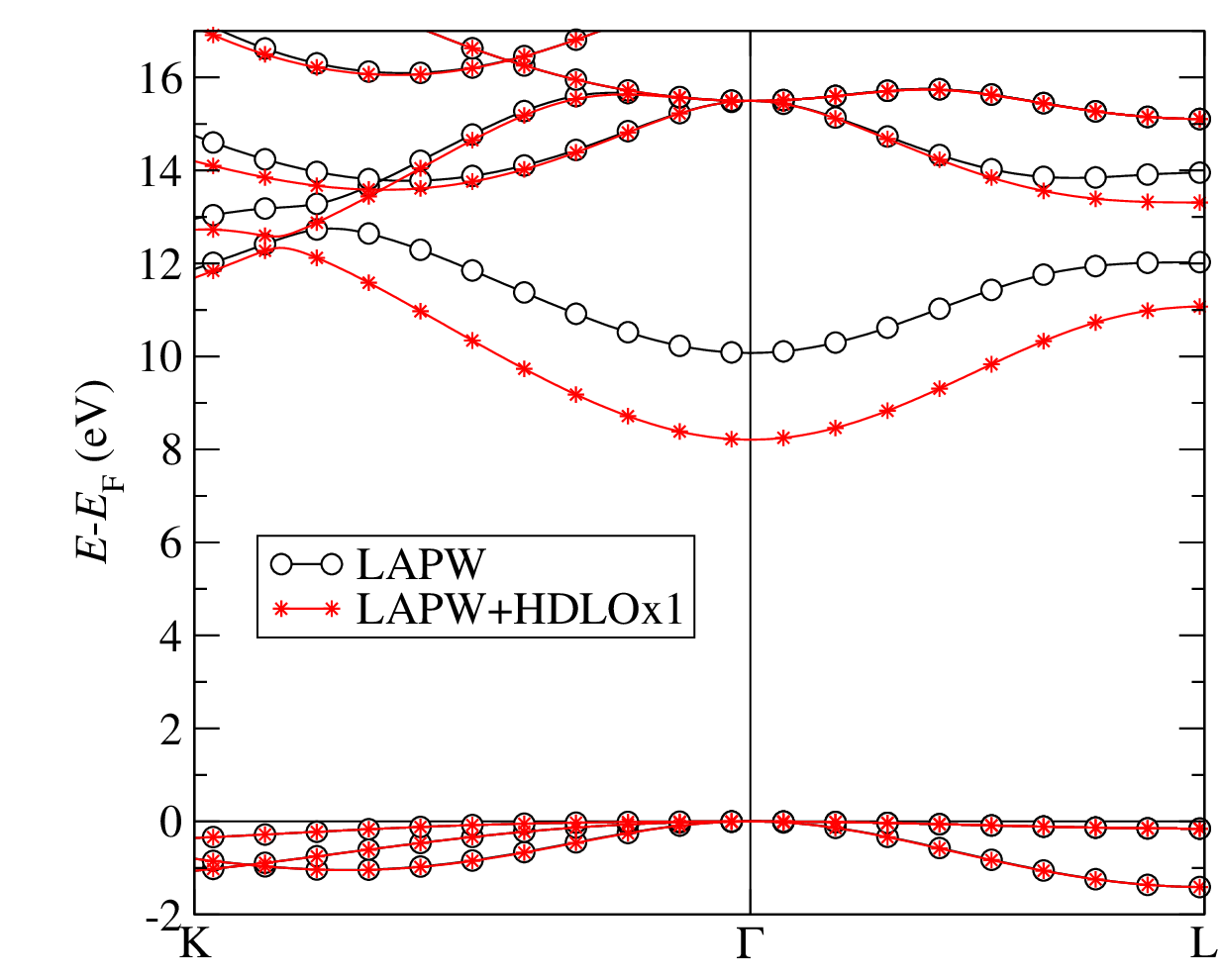}
\caption{Same as Fig.~\ref{bandStructureKClFigure} for Ar.}
\label{bandStructureArFigure}
\end{figure}

In Figs.~\ref{bandStructureKClFigure} and \ref{bandStructureArFigure}
the KS band structures of 
rock-salt KCl and crystalline Ar, respectively, are shown. Both exhibit a direct band gap at $\Gamma$.
We show band structures calculated with the conventional LAPW and 
the LAPW+HDLO$\times$1 basis (results obtained by the HELO extensions lie on
top of the LAPW+HDLO$\times$1 ones on the scale of the diagrams).
As expected, the occupied states, being close to the energy parameters, are
already well described in the conventional LAPW basis and the 
occupied bands are indistinguishable on the energy scale of the diagrams. 
However, there are clear deviations in the unoccupied states.
In particular, the lowest unoccupied band, 
which is of $4s$ character, shifts down by 0.19~eV for KCl and even by 1.87~eV
for Ar upon introducing the LOs; the band
dispersion is also affected noticeably. The downward 
shift of the lowest unoccupied band entails, of course, a reduction 
of the KS band gaps, which amounts to 4\% for KCl and 19\% for Ar.
As a result, the linearization error has an even more pronounced effect on
the band gap than it has on the total energy and equilibrium lattice constant.
Higher-lying unoccupied states will be affected even more strongly so that
a thorough convergence of the basis set is mandatory in the calculation of 
quantities that involve the unoccupied states in a perturbative way 
such as response quantities used,
e.g., in the optimized-effective-potential method and the $GW$ approximation
\cite{PhysRevB.83.045105,PhysRevB.83.081101}.

\section{Basis-set size}
\label{basisSetSize}

\begin{figure}
\centering
\includegraphics*[width=\columnwidth]{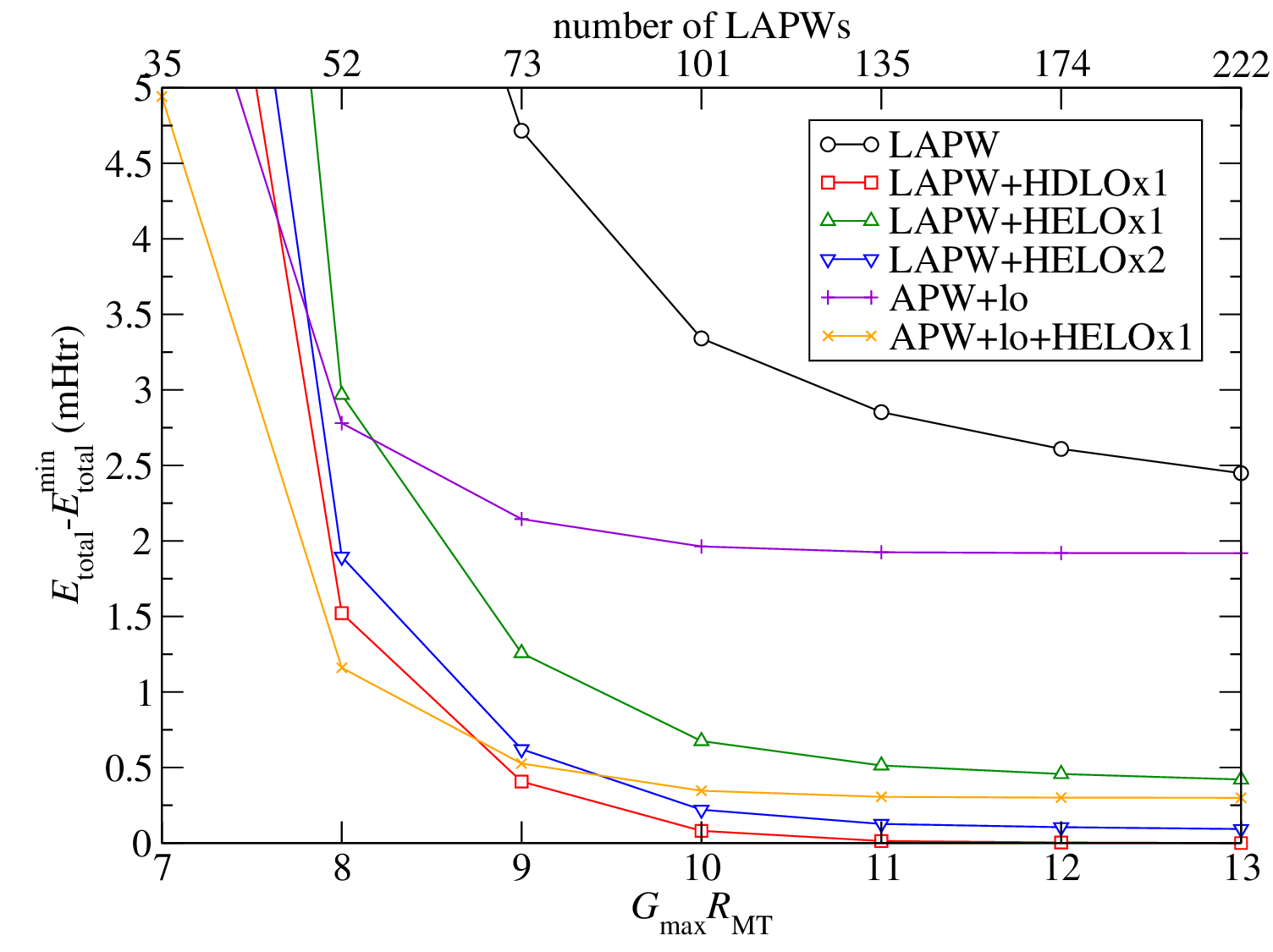}
\caption{Convergence of the total energy for fcc Ce with respect to
$G_{\mathrm{max}}R_{\mathrm{MT}}$ with
$R_{\mathrm{MT}}=3.14 \thinspace a_0$. The total energy is given
relative to the minimal energy obtained with the LAPW+HDLO$\times$1 basis.
The upper abscissa of the figure gives the size of the conventional LAPW
basis averaged over all $\vec{k}$ points. For basis sets with one set of LOs (dashed lines, 
LAPW+HELO$\times$1, LAPW+HDLO$\times$1, APW+lo) the total number of basis functions is 
increased by 16, for those with two sets (dotted lines, LAPW+HELO$\times$2,
APW+lo+HELO$\times$1) it is increased by 32. The best performance in terms of rate of
convergence and accuracy of the converged value is achieved with the
LAPW+HDLO$\times$1 basis.}
\label{convergenceKmaxRmt}
\end{figure}

In this section, we compare for the example of fcc Ce the speed of convergence of the total energy with respect to
the cutoff radius of reciprocal lattice vectors achieved with the different
basis sets. In Fig.~\ref{convergenceKmaxRmt} the convergence of the
total energy is shown as a function of $G_{\mathrm{max}}R_{\mathrm{MT}}$
with $R_{\mathrm{MT}} = 3.14 \thinspace a_0$. In addition to the
conventional LAPW basis and those extended by HDLOs and HELOs,
we also consider the convergence for the 
APW+lo basis and its HELO$\times$1 extension. In contrast to the LAPW basis, 
the APW+lo approach \cite{Sjoestedt200015} augments the interstitial plane
waves solely by the functions $u_l(r,E_l)Y_{lm}(\hat{\vec{r}})$,
while the energy derivatives $\dot{u}_{l}(r,E_l)$ up to a given angular momentum (here, $l=3$) are
treated as LOs (for which we have adopted the common lower case notation ``lo''). 
For higher $l$ we depart from the original APW+lo approach and follow the idea of 
Madsen \textit{et al.}~\cite{PhysRevB.64.195134} to use both radial functions for augmentation as in LAPW.
As a consequence of the altered linearization, the APW+lo basis is only continuous in value but
shows a discontinuity in the slope at the MT sphere boundary, i.e.,
the basis functions exhibit a kink there.
The less stringent matching at the MT sphere boundary
leads to a more flexible basis set, 
which manifests in a faster convergence of the total energy with respect 
to $G_{\mathrm{max}}R_{\mathrm{MT}}$ in comparison to the conventional LAPW basis. This effect is clearly visible in 
Fig.~\ref{convergenceKmaxRmt}. While the conventional LAPW basis does not 
yield a converged total energy even with $G_{\mathrm{max}}R_{\mathrm{MT}}=13$, 
the APW+lo curve reaches its converged value already for
$G_{\mathrm{max}}R_{\mathrm{MT}} \approx 10$. We note that the LAPW basis 
finally converges to the same total energy as the APW+lo approach, but at 
a considerably larger basis, even when taking account of the extra set of LOs in APW+lo. (The same holds for the APW+lo+HELOx1 and LAPW+HELOx1  curves.). This is in agreement with earlier investigations 
on fcc Ce performed by Sj\"ostedt \textit{et al.} \cite{Sjoestedt200015}.

However, if we extend the LAPW basis by one set of LOs (rather than increasing $G_\text{max}$), 
thus making the basis set as large as in the APW+lo method, we reach a lower and more 
accurate total energy than in the APW+lo method due to a larger variational freedom in the MT spheres.
Among the evaluated basis sets the LAPW+HDLO$\times$1 basis 
reaches the variationally lowest total energy. We note that adding further
HDLOs (i.e., the third and higher derivatives) hardly changes the converged value.

We can also observe that the corresponding calculations converge more 
rapidly than those employing the conventional LAPW basis. The reason for
that is the increased flexibility brought about by the additional LOs. This
increased flexibility plays a similar role as allowing for a kink in the
APW+lo basis functions. So, it acts toward a decoupling of the two regions of
space, allowing to some extent a separate variational search for the 
eigenstates in the two regions. The decoupling is somewhat less effective in the LAPW 
basis sets, where the radial derivatives of the wave functions are forced to be continuous 
at the MT sphere boundaries. However, it should be pointed 
out that every kinkless wave function that is given as 
a linear combination of plane waves in the IR and the $u_{l}(r,E_l)$ and $\dot{u}_{l}(r,E_l)$
in the MT spheres is representable by the LAPW basis functions, which are constructed to fulfill 
exactly these matching conditions. Hence, 
whenever an APW+lo calculation yields a smaller total energy than the corresponding LAPW
calculation, then the wave functions necessarily exhibit a kink at the MT
sphere boundaries, which from a physical point of view might be unsatisfactory.
Loosely speaking, the variational search for the wave functions reached a
lower total energy by accepting a kink with a singular kinetic energy density at the sphere
boundaries.

\section{Conclusion}
\label{conclusion}

In foresight of new challenges treating solids with greater electronic and chemical complexity with more 
sophisticated functionals, we have explored in this work the capability of the LAPW basis set to deal with these challenges
and evaluated extensions of the basis set by two types of local orbitals that basically offer the potential to 
provide the density functional answer to the problem at hand with very high precision. 

In detail we have analyzed the effects of the linearization error
that arises in the LAPW basis due to the restriction to only two radial functions,
$u_l(r,E_l)$ and $\dot{u}_l(r,E_l)$, per $l$ quantum number in the MT spheres.
While the LAPW basis is established to be very accurate for a wide range of 
materials and material properties, there are cases where the linearization error
causes the results to depend appreciably
on numerical parameters that are inherent to the FLAPW method and are not convergence parameters, i.e.,
the energy parameters and the MT radii. Although choosing small MT radii
reduces the linearization error, this extends the IR and thus entails a 
larger reciprocal cutoff radius leading to many additional basis functions.
A much more effective way is to add LOs, for example HELOs, which are defined with 
higher energy parameters. HELOs are often used to improve the description of unoccupied states, 
but can also be employed to greatly reduce the linearization error for the occupied states. An even 
more efficient way to eliminate the linearization error for the occupied states, and thereby to
make the results insensitive to variations of the energy parameters and MT radii, is to extend the 
basis with HDLOs, which are constructed from the second energy derivatives $\ddot{u}_l(r,E_l)$.

The addition of LOs allows to employ large MT spheres,
which keeps the reciprocal cutoff radius small and thus keeps the basis-set size to a
minimum. Using large spheres also ensures that the orthogonality of the
basis functions for the valence states to the core states is maximized,
which avoids ghost bands to appear in the valence and conduction band structure. The inclusion 
of extra sets of LOs might not always be necessary, but in the light of the present results we 
recommend it to be a routine part in the convergence of FLAPW calculations with respect to the basis 
set, in particular in systems with large valence band widths and if large MT radii are employed. We 
have found that a single set of HDLOs (16 additional functions per atom) already improves the 
accuracy of the results by one or even two orders of magnitude, yielding sufficient accuracy for all 
practical purposes. On the other hand, HELOs are very effective in improving the description of 
high-lying conduction states. This is of importance when Kohn-Sham wave functions are used as 
input for more sophisticated methods, such as the $GW$ approximation or the total energy in the 
random-phase approximation.

Beyond the efficient elimination of the linearization error, the flexibility of the basis at the MT sphere boundaries is a major aspect controlling the required basis-set size. Extending the conventional LAPW basis set by LOs helps to decouple the two regions of space (MT spheres and interestitial region) in a similar way as in the related APW+lo approach resulting in a faster basis-set convergence. However, the decoupling is somewhat less effective than in APW+lo, where the matching condition of the radial derivative is dropped completely admitting single-particle wave functions that exhibit a kink at the MT sphere boundary. However, the LAPW+LO basis---here, HDLOs are most efficient---contains one radial function more than the APW+lo method so that a higher precision is achieved at equal basis-set sizes, in cases where the linearization error is large.

\section{Acknowledgment}
We thank Eugene Krasowskii for a careful reading of the manuscript.



%% References


\bibliographystyle{elsarticle-num}
\bibliography{references.bib}






% \end{thebibliography}

\end{document}